\documentclass[prb,aps,reprint,showpacs,twocolumn]{revtex4-1}
\usepackage{graphicx,amsmath}
\usepackage{float}
\usepackage{subfigure}
\usepackage{mathrsfs}
\usepackage{graphics}
\usepackage{epstopdf}
\usepackage{amsmath}
\usepackage{amssymb}
\usepackage{hyperref}
\usepackage{color}
\usepackage{braket}
\usepackage{multirow}
\usepackage[normalem]{ulem}
\begin{document}
\author{Kallol Mondal and Charudatt Kadolkar}
\affiliation{Department of Physics, Indian Institute of Technology Guwahati, Guwahati, Assam 781039, India}
\date{\today}
\title{$\mathbf{Q=0}$ order in quantum kagome Heisenberg antiferromagnet}
\begin{abstract}
We have studied the nearest neighbor Heisenberg model with added Dzyaloshinskii-Moriya interaction using Schwinger boson mean-field theory considering the in-plane component as well as out-of-plane component. Motivated by the experimental result of vesignieite that the ground state is in a $\mathbf{Q=0}$ long-range order state, we first looked at the classical ground state of the model and considered the mean-field \textit{ansatz} which mimics the classical ground state in the large $S$ limit. We have obtained the ground-state phase diagram of this model and calculated properties of different phases. We have also studied the above model numerically using exact diagonalization up to a system size $N=30$. We have compared the obtained results from these two approaches. Our results are in agreement with the experimental result of the vesignieite.
\end{abstract}
\maketitle
\section{\label{sec:intro}Introduction}
Geometrically frustrated magnets are the potential candidate to host exotic ground states like quantum spin liquids, a state with fractional excitations, high entanglement, and without any broken symmetries even at $T=0$~\cite{anderson1973resonating,savary2016quantum,balents2010spin,zhou2017quantum}. The most promising candidate to possess spin liquid ground state is the spin 1/2 kagome lattice with vertex sharing triangles. Antiferromagnetic ordering of the spins on a kagome lattice is frustrated by the very nature of the geometry of the lattice. The key features which make it suitable for the spin liquid ground state are its low dimensionality and a higher degree of frustration. The low value of spin leads to the large quantum fluctuation, hence preventing the ordered ground state~\cite{PhysRevB.47.5459,yan2011spin,PhysRevLett.101.117203,PhysRevLett.98.117205}. However, in search of material realizing quantum spin liquid ground state, materials tend to deviate from the perfect kagome structure due to the presence of disorder, structural distortion, Dzyaloshinskii-Moriya interaction(DMI) or other long-range interactions. However, a theoretical study reveals that the presence of minute perturbation may have a deep impact on the ground state manifold~\cite{elhajal2002symmetry}. One such perturbation that is very sensitive to the low-temperature magnetic structure of this frustrated magnets is Dzyaloshinskii-Moriya interaction~\cite{moriya1960anisotropic,dzyaloshinskii1957thermodynamic}. DMI appears in a lattice where there is a lack of inversion symmetry between the two magnetic sites, was first introduced to explain the weak ferromagnetism in $\alpha$-Fe$_2$O$_3$ ~\cite{dzyaloshinskii1957thermodynamic}. The interaction term is of the form $H^\prime_{ij} = \vec{\mathbf{D}}_{ij} \cdot (\vec{\mathbf{S}}_i \times \vec{\mathbf{S}}_j) $ where $\vec{\mathbf{D}}_{ij}$ is the Dzyaloshinskii-Moriya (DM) vector, the strength of the coupling and $i,j$ are the site index. The DM vector $\vec{\mathbf{D}}_{ij}$ lies on a mirror plane bisecting the bond joining the two magnetic sites $i$ and $j$

In the experimental side, several materials were thought to be potential candidates for the kagome antiferromagnet (KAFM) to host a quantum spin liquid ground state like herbertsmithite~\cite{shores2005structurally},volborthite~\cite{hiroi2001spin} and vesignieite~\cite{okamoto2009vesignieite}. Among all these, the mineral Herbertsmithite is found to be a geometrically perfect description of quantum kagome Heisenberg antiferromagnet(QKHAF), which has been studied intensively~\cite{mendels2016quantum}. This material does not show any sign of ordering down to 50 mk, which is 3000 times lower than the characteristic exchange energy. Herbertsmithite is strongly suspected of hosting a quantum spin liquid ground state with spinon excitations.~\cite{helton2007spin,lee2008end} Form ESR data, the measured value of the in-plane component of DMI comes out as $0.01J$, whereas the out-of-plane of DMI is much larger $0.06J$~\cite{zorko2008dzyaloshinsky,PhysRevB.81.224421}. Exact diagonal results predict that there may be quantum critical point $D_c = 0.1 J$, at the one side $D < D_c $ there is moment free phase and on the other side $D > D_c$ there is Neel ordered phase.~\cite{cepas2008quantum}

Unlike herbertsmithite, vesignieite shows a magnetic transition to $\mathbf{Q=0}$ magnetic order with the in-plane moments on the three sublattices oriented at $120^{o}$ with each other, at a surprisingly high temperature $T_N = 9 K $.~\cite{okamoto2009vesignieite,PhysRevB.88.144419,yoshida2012magnetic} ESR spectra reveal the presence of large DMI anistropy~\cite{zhang2010high}. So, it is expected that a large value of DMI may lead to the $\mathbf{Q=0}$ magnetic structure i.e., the other side of the quantum critical point~\cite{quilliam2011ground}. In contrast to herbertsmithite, the dominant anisotropy is the in-plane component of DMI. For vesignieite, the measured value of the in-plane component is found to be $0.19J$, and the out-of-plane component is $0.07J$, as indicated by ESR data analysis.

Apart from herbertsmithite, vesignieite, the compound $\text{Nd}_3\text{Sb}_3\text{Mg}_2\text{O}_{12}$ is also of much interest since it shows large canting angle $\eta = 30.6^o$ indicating the presence of large in-plane DMI, $D_p = 0.8 J$ as predicted by Scheie et. al. ~\cite{PhysRevB.93.180407}. However Laurell et. al. ~\cite{PhysRevB.98.094419} argued that the predicted value should be $D_p > 1.5 $ to reproduce the such large canting angle. There are also interesting cases with even larger DMI as precdicted from first principle calculation~\cite{PhysRevMaterials.2.074408}.
\begin{figure*}[ht!]
\includegraphics[height=5.75cm,width=8.5cm]{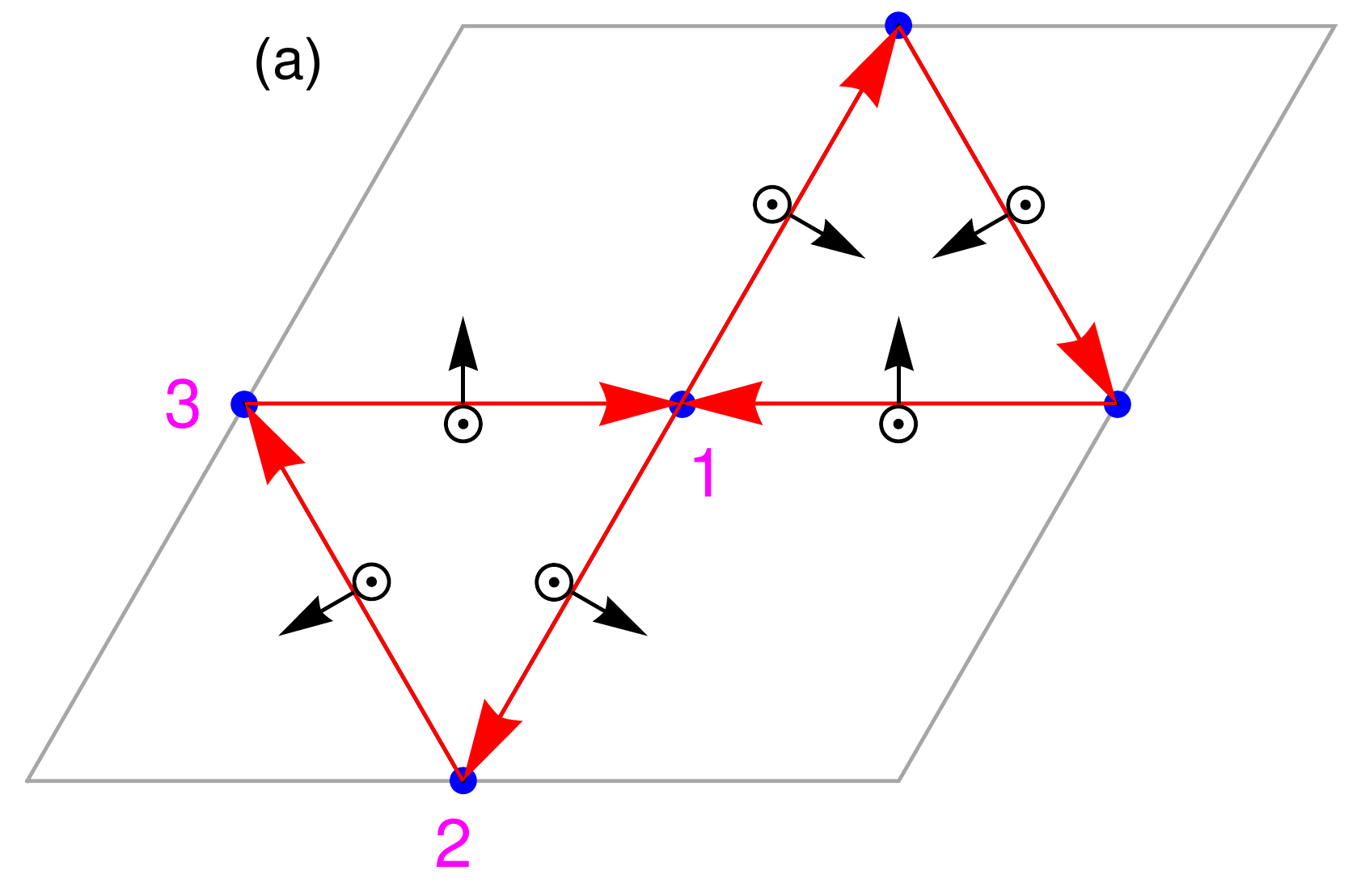}\hspace{0.75cm}
\includegraphics[height=6cm,width=7cm]{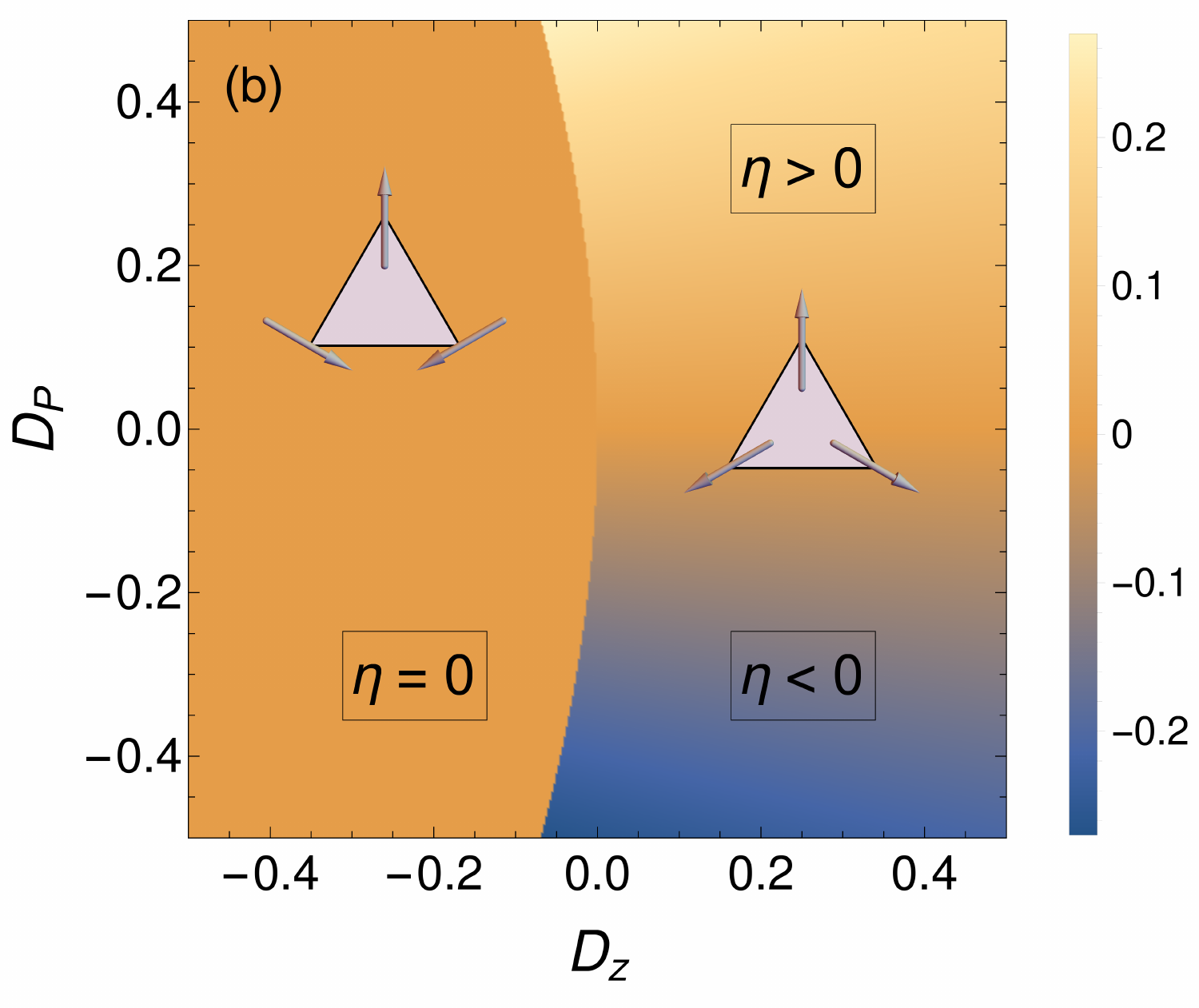}
\caption{(a) The orientation of DM vector, in-plane component $D_p$ is shown by the black arrow and the out-of-plane component $D_z$ is uniform along $\hat{z}$. (b) Classical ground state phase diagram for spin-1/2}
\label{DM-config-classical-pd}
\end{figure*}

In this work, we study the ground state of the nearest neighbor Heisenberg model with added Dzyaloshinskii-Moriya interaction using the Schwinger boson mean-field theory(SBMFT) framework and numerically using exact diagonalization(ED) method up to cluster size $N=30$. There are several SBMFT studies as well as an exact diagonalization study, which only focuses on the out-of-plane component od DMI~\cite{messio2010schwinger,halimeh2016spin,PhysRevB.95.134404,PhysRevLett.118.267201,cepas2008quantum}. In the present study, we consider the in-plane and the out-of-plane component of DMI both to study the ground-state phase diagram. We have compared the results obtained from these two different approaches.

The layout of this paper is as follows. In sec. II, we discuss the model Hamiltonian and the orientation of the DM vector. In sec. III, we briefly describe the classical ground state of this model. In sec. IV, we present the Schwinger boson formalism. In sec. IV, we present the result obtained from the SBMFT approach. In sec. V, we discuss the exact diagonalization results of the proposed model. We compare the results obtained from these two distinct approaches and discuss their relevance in the experiment in sec. VI. Finally, in sec. VII, we make the concluding remarks.
\section{Model Hamiltonian}
In this work, we have explored the model Hamiltonian of vesignieite, as obtained in the Ref.~\cite{PhysRevB.88.144419}. For vesignieite, they found that the strength of  symmetric anisotropic exchange(AE) is comparable to the Dzyaloshinskii-Moriya interaction term and argued that since DMI results form the first order correction of $J$  in the spin-orbit coupling where as AE is the second order correction. Naturally DMI is supposed to be more influential on the low temperature magnetic structure~\cite{PhysRevB.88.144419}. So, the effective spin Hamiltonian for vesignieite  is given by
\begin{equation}
H= \sum_{\langle ij \rangle}  [ J_{ij} \vec{\mathbf{S}}_i \cdot \vec{\mathbf{S}}_j + \vec{\mathbf{D}}_{ij} \cdot  (\vec{\mathbf{S}}_i \times \vec{\mathbf{S}}_j)  ]
\label{eqn1:model}
\end{equation}
where the isotropic exchange interaction strength $J_{ij} = J$ for the nearest-neighbour pairs. The Dzyaloshinskii-Moriya vector, $\vec{\mathbf{D}}_{ij}$ has $D_p$ and $D_z$ as the strengths of the in-plane and  out-of-plane components of DMI. $\langle ij \rangle$ indicates the interactions are restricted to nearest neighbor only.   The order of the cross product between $i$-th and $j$- th site for given $\vec{\mathbf{D}}_{ij}$ are denoted by arrows as shown in the Fig.~\ref{DM-config-classical-pd}(a) .The lattice vectors are $\vec{a} = a (1,0)$ and $\vec{b} = a (\frac{1}{2},\frac{\sqrt{3}}{2})$. The DM vector is given by
\begin{subequations}
\begin{eqnarray}
\vec{\mathbf{D}}_{31}  & = & D_p \hat{j} + D_z \hat{k}  \\
\vec{\mathbf{D}}_{12}  & = & \hat{R}(\hat{k},-\frac{2 \pi}{3}) \vec{\mathbf{D}}_{31} \\
\vec{\mathbf{D}}_{23}  & = & \hat{R}(\hat{k},-\frac{4 \pi}{3}) \vec{\mathbf{D}}_{31}
\end{eqnarray}
\end{subequations}
where $\hat{R}(\hat{k},\theta)$ is the rotation operator that rotates a vector by an angle  $\theta$ about the  axis $\hat{k}$. The orientation of the out-of-plane and in-plane components of DMI are shown in the Fig.~\ref{DM-config-classical-pd}(a).

The introduction of DM interaction reduces the symmetries of the isotropic Heisenberg model. When $\vec{\mathbf{D}}_{ij} = D_z \hat{\mathbf{k}}$, that is when $D_p=0$, then the global spin rotation symmetry reduces to $U(1)$ from $SU(2)$ but the wallpaper group remains $p6m$. When $D_p \ne 0$, then there are global spin rotation symmetries, and the wallpaper group reduces to $p3m1$.

\begin{figure*}[ht!]
\includegraphics[height=6cm,width=6cm]{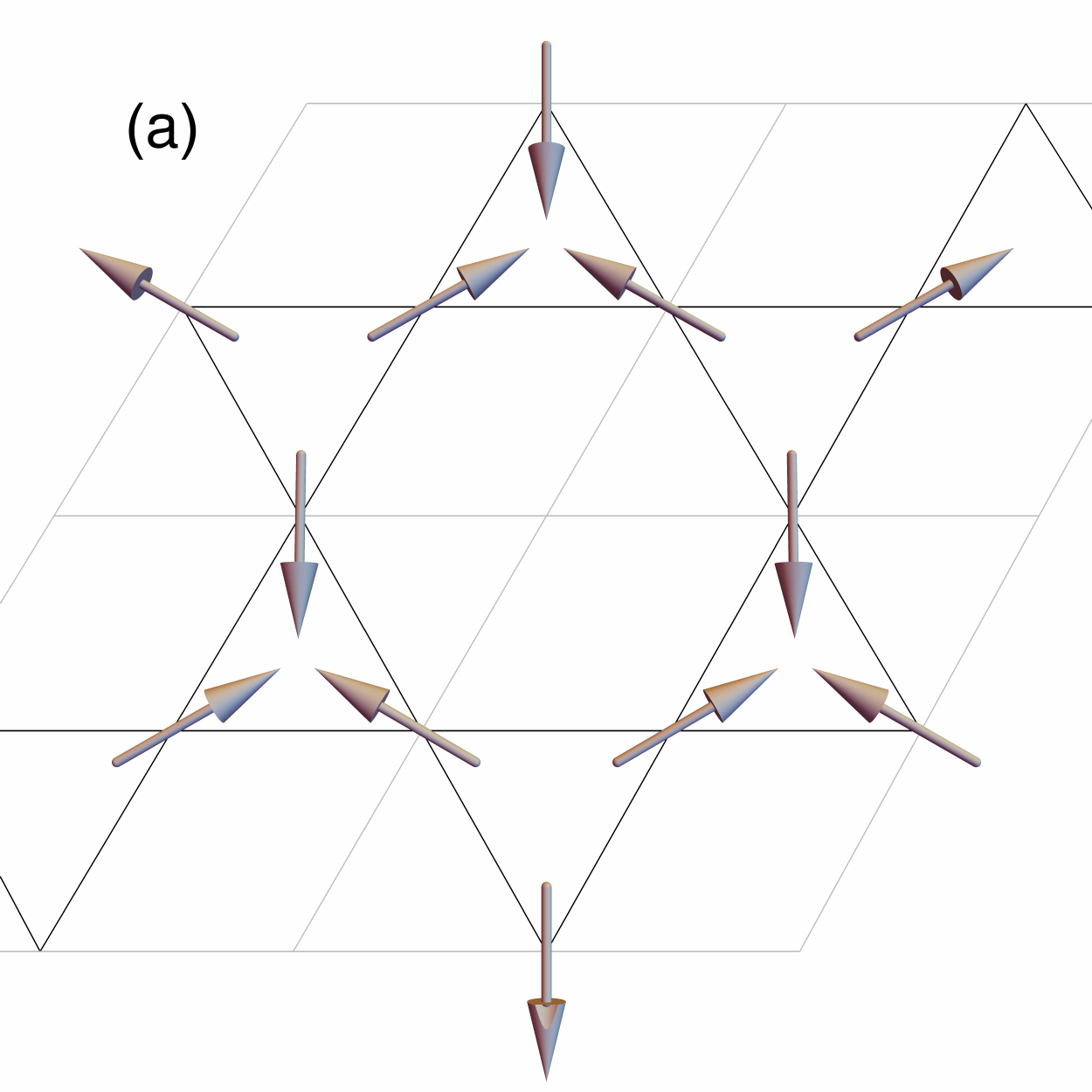}\hspace{0.3cm}
\includegraphics[height=2.5cm,width=2cm]{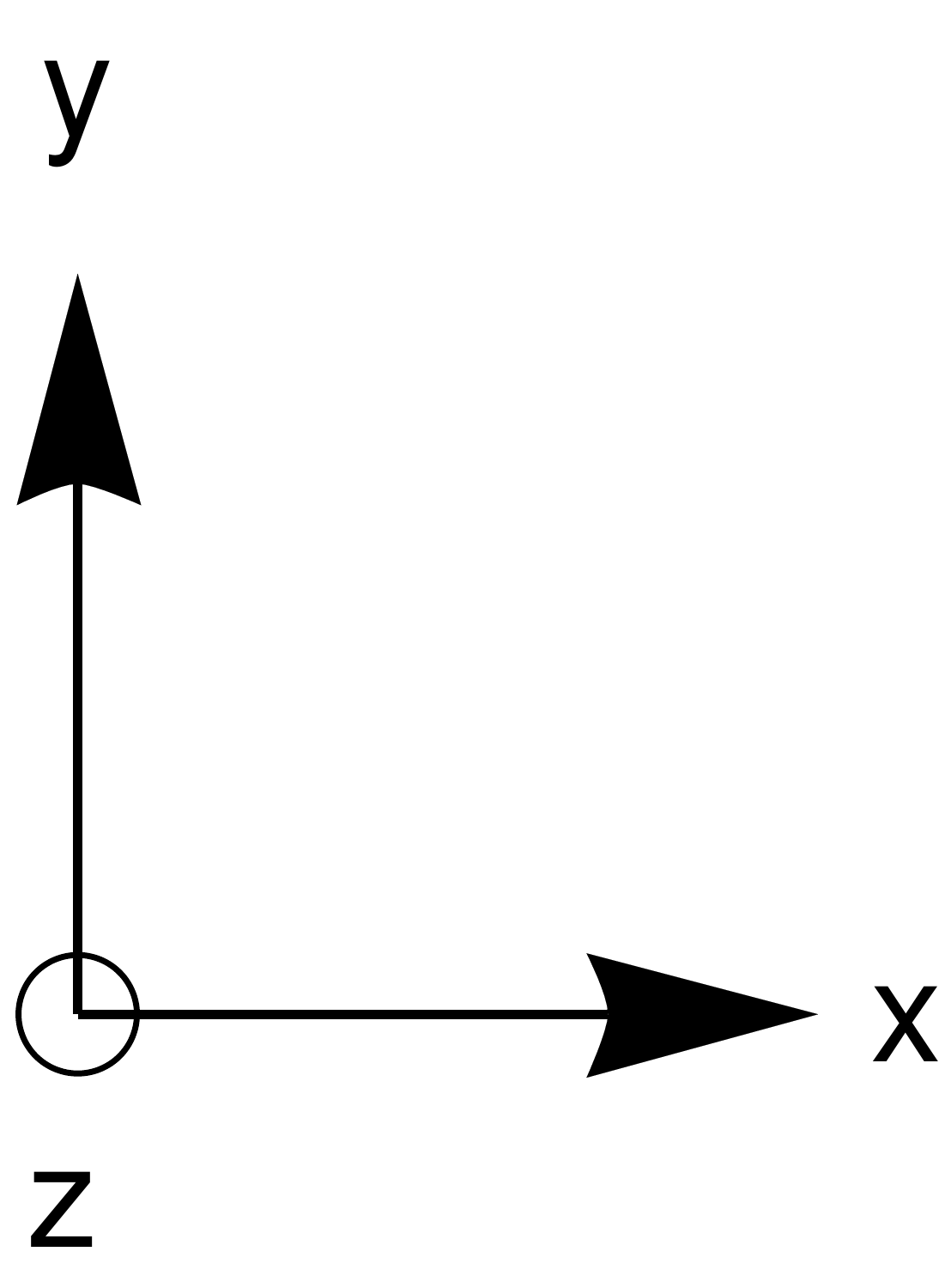}\hspace{0.5cm}
\includegraphics[height=6cm,width=6cm]{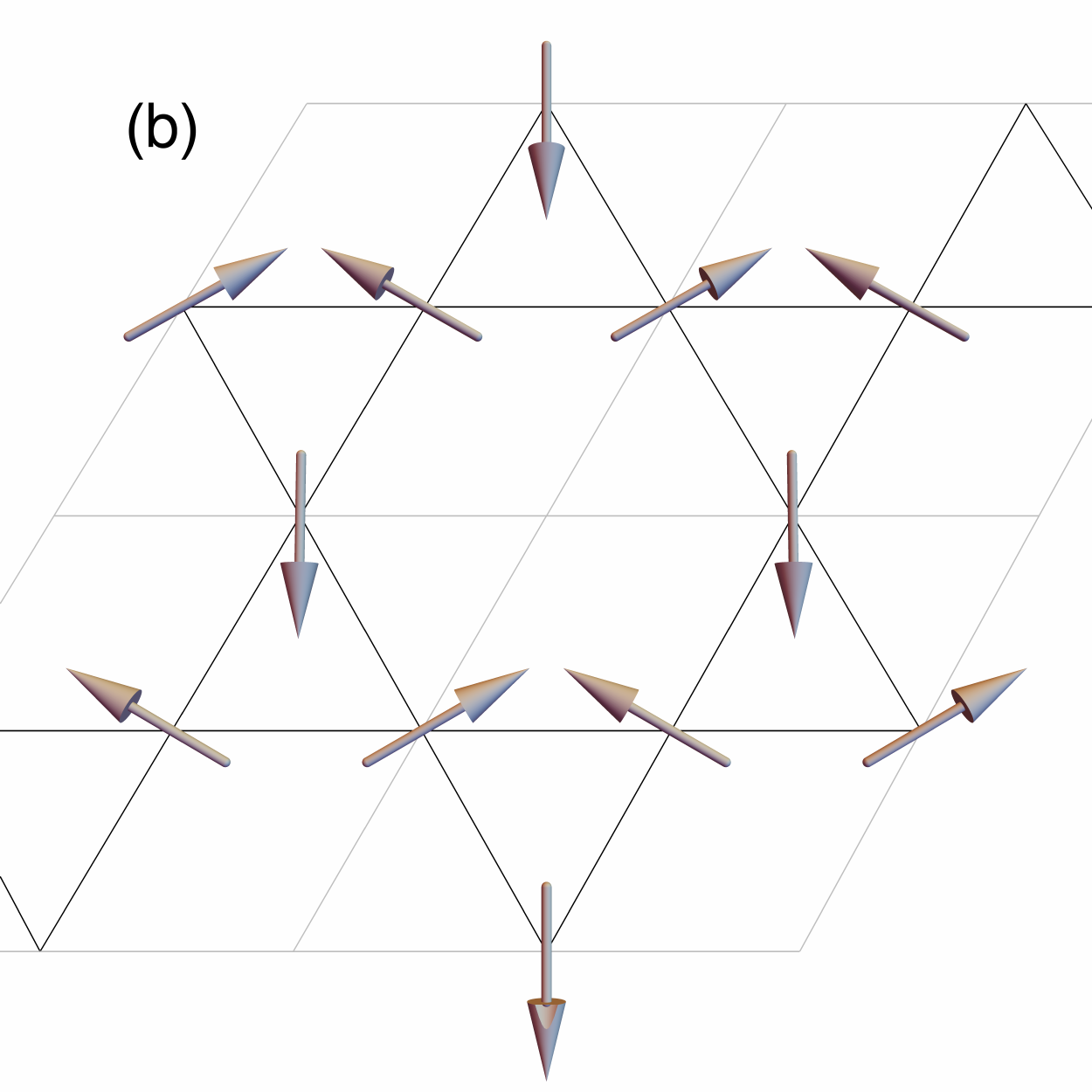}
\caption{(a) Positive chirality as  $D_z > 0$ and (b)  Negative chiralty as $D_z < 0$ }
\label{fig:chirality}
\end{figure*}

The classical ground state was discussed by Elhajal et al.~\cite{elhajal2002symmetry}. We introduce their results since it is relevant to the SBMFT used in the later sections. First, consider the pure isotropic case i.e., the absence of Dzyaloshinskii-Moriya interaction. Based on the projective symmetry group(PSG) analysis, Messio et al.~\citep{messio2011lattice} showed that there is eight possible classical magnetic structure in a kagome lattice, termed as regular magnetic order(RMO). They suggested that these states can be good variational candidates to compute the ground-state phase diagram in the mean-field approach. The states are given by (i) Ferromagnetic state (ii) $\mathbf{Q =0} $ state (iii) $\sqrt{3} \times \sqrt{3}$ states (iv) octahedral states (v) cuboc1 state (vi) cuboc2 state (vii) $\mathbf{Q =0}$ umbrella state and (viii) $\sqrt{3} \times \sqrt{3}$ umbrella states. Classical energies and the structure factor for these states can be found in Ref.~\cite{messio2011lattice}.

Motivated by the experimental result of vesignieite that there is a strong presence of DMI, and the
ground state is found to be $\mathbf{Q =0}$ long-range order(LRO) state. In the following, we discuss the $\mathbf{Q =0}$ classical
ground state of kagome Heisenberg antiferromagnet with the Dzyaloshinskii-Moriya interaction. The
classical ground state of kagome Heisenberg antiferromagnet is highly degenerate. All the possible
states, where the three spins respect the angle $2 \pi/3$ with each other to minimize the ground
state energy. With the introduction of the DMI, the symmetry is lowered, though the ground state is
non-coplanar $\mathbf{Q=0}$ LRO with planar components of the spins making $2\pi/3$ angle with each other.

With lowered symmetries, the spin arrangements can now be classified using the notion of chirality,
that is the angle
between the spins in a given triangle of Kagome lattice. If $\vec{S}_{1},\vec{S}_{2}$ and $\vec{S}_{3}$
are three spins in a given triangle located in counter-clockwise direction, then we define the spin chirality as
\begin{equation}
\chi_z =
\left[
\vec{\mathbf{S}}_{1} \times \vec{\mathbf{S}}_{2} + \vec{\mathbf{S}}_{2} \times \vec{\mathbf{S}}_{3} +\vec{\mathbf{S}}_{3} \times \vec{\mathbf{S}}_{1}
\right]\cdot\hat{\mathbf{k}}
\label{eqn:chirality}
\end{equation}
In addition, the spins are non-colplanar, all the spins make a canting angle $\eta$ with the plane of the lattice.
The energies for these spin configurations are given by
\begin{eqnarray}
\frac{E_{+}}{N} & = & \frac{J}{2}[1- 3 \cos(2 \eta)]-\sqrt{3} (D_z \cos^2\eta +D_p \sin(2\eta)\cos\phi)\nonumber \\
\frac{E_{-}}{N}& = & \frac{J}{2}[1- 3 \cos(2 \eta)]+\sqrt{3} D_z \cos^2\eta
\label{eqn:classical-energy}
\end{eqnarray}
where $N$ is the total number of spins and  $E_+(E_-)$ is the energy of the configuration(as shown in the Fig.~\ref{fig:chirality}) with positive(negative) chirality, and $\phi$ is the angle azimuthal
angle of $\vec{S}_1$.

From the energy expressions, we can describe the ground state spin configuration. In the absence of in-plane
component, the Hamiltonian is invariant under rotation around the z-axis. The spins are forced to lie in the lattice plane with positive or negative chirality depending on the sign of $D_z$. In a more interesting case, when both in-plane and out-of-plane are present; there are two phases distinguished by chirality. For large negative $D_z$, the spins remain coplanar with negative chirality irrespective of the value of $D_p$. For positive $D_z$, the spins have positive chirality but make a canting angle
\begin{equation}
\eta = \frac{1}{2}\tan^{-1} \Big[\frac{2 D_p}{\sqrt{3} J + D_z}\Big]
\label{eqn:canting-angle}
\end{equation}
giving rise to weak ferromagnetism.

The complete phase diagram is shown in Fig.~\ref{DM-config-classical-pd}(b). The two phases are separated by a first-order transition. For negative $D_z$, the canting angle continues to grow with the value of $D_p$. In the phase diagram, the canting angle is shown using a color map.

\section{Schwinger Boson formalism}
One of the advantages of SBMFT formalism is that this approach can address both the long-range ordered states and the spin liquid states. Unlike fermionic approach long-range order appears due to the condensation of the Schwinger bosons and hence the liquid spin states
will have gapped bosonic spinons.

The model Hamiltonian Eq.~\ref{eqn1:model} can be mapped to a simpler model with $U(1)$ symmetry up to terms second order in $D_p$. This is due to the fact that the vector sum of the in-plane components of DMI is zero in a triangle. So, for small values of in-plane components of DMI, we can rotate the spins in such a way that the $U(1)$ symmetry is restored~\citep{cepas2008quantum}. However, when the strength of the in-plane component of DMI is comparable to or greater than the strength of the out-of-plane component of DMI, then the above rotation will not work. 

To treat this problem, we use the method employed by Manuel et al~\cite{Manuel1996}.  Consider one bond between site $i$ and site $j$. The Hamiltonian for this bond is 
\begin{equation}
	H_{ij} =   J \left[ \vec{\mathbf{S}}_i \cdot \vec{\mathbf{S}}_j + 2 \tan\left(\theta\right) \hat{\mathbf{d}}_{ij} \cdot  			(\vec{\mathbf{S}}_i \times \vec{\mathbf{S}}_j) \right]
\end{equation}
where $\tan\theta = \left| \vec{\mathbf{D}}_{ij} \right|/ 2 J = \sqrt{D_p^2 + D_z^2}/2 J$ and $\hat{\mathbf{d}}_{ij} $ is the unit vector along $\vec{\mathbf{D}}_{ij}$. Now, we rotate the spins at $i$-th and $j$-th sites by the angles $\theta_i=\theta$ and $\theta_j=-\theta$ about the axis $\hat{\mathbf{d}}_{ij}$. Under this rotation, let $\mathbf{S}_i\mapsto\mathbf{S}_i^\prime$ and $\mathbf{S}_j\mapsto\mathbf{S}_j^\prime$. Using the fact that $\theta$ is small and neglecting terms in $\theta^2$, we can show that the bond Hamiltonian becomes (see Appendix for the derivation)
\begin{equation}
H_{ij} = J \vec{\mathbf{S}}^\prime_i \cdot \vec{\mathbf{S}}^\prime_j.
\label{eqn:rotated-H}
\end{equation}
Note that $\vec{\mathbf{S}}^\prime_i$ that appears in two different bond Hamiltonians, say $\mathbf{H}_{ij}$  and $\mathbf{H}_{ik}$, are not the same since the rotation axes $\hat{\mathbf{d}}_{ij}$ and $\hat{\mathbf{d}}_{ik}$ are different.

\vspace{0.5cm}
In Schwinger boson formalism, the spin operator is represented by two bosonic operators  $ a$ and $ b$, given by
\begin{equation}
\mathbf{S} = \mathbf{\Phi}^\dagger \cdot \mathbf{\sigma} \cdot \mathbf{\Phi}
\end{equation}
with $\mathbf{\Phi} \equiv (a,b)$ be the bosonic spinor and $\mathbf{\sigma} $ be the vector of Pauli matrices.
 The component of spin along an arbitrary direction $\hat{n}$ is given by
\begin{eqnarray}
\mathbf{S}_{\hat{n}} \mapsto \frac{1}{2} \mathbf{\Phi}^\dagger  ( \hat{n}\cdot \sigma) \mathbf{\Phi} \nonumber
\end{eqnarray}  
In Schwinger boson formalism there is a $U(1)$ gauge symmetry as $a_i \rightarrow e^{i \phi(i)} a_i$ and $b_i \rightarrow e^{i \phi(i)} b_i$. The boson operators obey the typical bosonic commutation relations $[  \mathbf{\Phi}_{i\alpha},  \mathbf{\Phi}^\dagger_{j \beta} ]=\delta_{ij} \delta_{\alpha \beta}$~\cite{Manuel1996}. This representation enlarges the Hilbert space. So, to remain within the physical space, the total number of Schwinger boson at a particular is constrained to be $2S$.  In the standard mean field treatment the constraint is implemented  by taking the ground state average and a Lagranges multiplier  $\lambda$ is introduced which can be thought as  chemical potential. Now we define two bond operator in the following way
\begin{equation}
A_{ij} = \frac{i}{2} \mathbf{\Phi}_i^{T} \sigma_y  \mathbf{\Phi}_j \hspace{0.25cm} \text{and} \hspace{0.25cm} B_{ij}^\dagger = \frac{1}{2}  \mathbf{\Phi}_i^{\dagger} \mathbf{\Phi}_j
\label{eqn:definitionAandB}
\end{equation}
The bond operator $A^\dagger_{ij} $creates a singlet at the bond $ij$ where as $B_{ij} $ helps the Schwinger boson to hop from site $i$ to site $j$. Since the form  of Hamiltonian in Eq.~\ref{eqn:rotated-H} is invariant under global spin rotation, we can always decouple the Hamiltonian in terms of two bond operator as given by
\begin{equation}
\mathbf{S}^\prime_i \cdot\mathbf{S}^\prime_j = : B^{\prime \dagger}_{ij}B^\prime_{ij}: -  A^{\prime \dagger}_{ij}A^\prime_{ij} 
\end{equation}
$::$ indicates the normal ordered product. The rotated bond operators can be written in terms of unrotated bond operators as following
\begin{subequations}
 \begin{eqnarray}
 B^{\prime \dagger}_{ij}& = & \cos\theta ~B^\dagger_{ij} + \sin\theta ~C^\dagger_{ij}  \\
 A^{\prime \dagger}_{ij} & = & \cos\theta ~A^\dagger_{ij} - \sin\theta ~D^\dagger_{ij}
 \end{eqnarray}
 \end{subequations}
 with two additional bond operators, given by
\begin{subequations}
\begin{eqnarray}
C^\dagger_{ij} & = & \frac{1}{2} \mathbf{\Phi}^\dagger_i (i ~\hat{\mathbf{d}}_{ij} \cdot \sigma ) \mathbf{\Phi}_j  \\
D^\dagger_{ij} & = & \frac{1}{2}  \mathbf{\Phi}^T_i (\sigma_y ~\hat{\mathbf{d}}_{ij} \cdot \sigma ) \mathbf{\Phi}_j 
\end{eqnarray}
\label{eqn:definitionCandD}
 \end{subequations}
With this bond operators we can identify the following identities
\begin{eqnarray}
\hat{\mathbf{d}}_{ij} \cdot   (\vec{\mathbf{S}}_i \times \vec{\mathbf{S}}_j)  =   \frac{1}{2} (:B^\dagger_{ij} C_{ij}&+& C^\dagger_{ij} B_{ij} : + A^\dagger_{ij} D_{ij} + D^\dagger_{ij} A_{ij})\nonumber \\
2 (\hat{\mathbf{d}}_{ij} \cdot \vec{\mathbf{S}}_i)  (\hat{\mathbf{d}}_{ij} \cdot \vec{\mathbf{S}}_j)   - \vec{\mathbf{S}}_i \cdot \vec{\mathbf{S}}_j &  = &   (:C^\dagger_{ij} C_{ij}: -D^\dagger_{ij}D_{ij}) \nonumber
\end{eqnarray}
The model Hamiltonian in terms of bond operators can be cast in the following way
\begin{equation}
H= \sum_{\langle ij \rangle }J ( \hat{B}^{\prime \dagger}_{ij} \hat{B}^{\prime }_{ij} -  \hat{A}^{\prime \dagger}_{ij} \hat{A}^{\prime }_{ij} )
\end{equation}
We can decouple the quadratic field Hamiltonian in terms of bilinear  operators using standard mean field decoupling scheme. The form of the mean field Hamiltonian is as following
\begin{equation}
H_{\text{MF}} = \sum_{\langle ij \rangle }J \Big[( \hat{B}^{\prime \dagger}_{ij} \mathbf{B}^{\prime }_{ij} -  \hat{A}^{\prime \dagger}_{ij} \mathbf{A}^{\prime }_{ij} ) + \text{H. C} \Big] - \sum_i \lambda_i \hat{n}_i  + \epsilon_0 \nonumber
\end{equation}
with the mean fields  corresponding to the bond operators is given by 
\begin{eqnarray}
 \langle \hat{B}^{\prime }_{ij} \rangle   =   \mathbf{B}^{\prime }_{ij} \hspace{0.5cm}  \text{and}  \hspace{0.5cm} \langle \hat{B}^{\prime \dagger }_{ij} \rangle =  \mathbf{B}^{ \prime * }_{ij} \nonumber \\
 \langle \hat{A}^{\prime }_{ij} \rangle =  \mathbf{A}^{\prime }_{ij} \hspace{0.5cm}  \text{and}  \hspace{0.5cm} \langle \hat{A}^{\prime \dagger }_{ij} \rangle =  \mathbf{A}^{ \prime * }_{ij} \nonumber\nonumber 
\end{eqnarray}
 with  $\epsilon_0$ is constant which depends on the mean fields and $\lambda$ given by  $\epsilon_0 = \sum_{\langle ij \rangle } [ | \mathbf{A}^{\prime }_{ij}|^2 - | \mathbf{B}^{\prime }_{ij}|^2 + 2 S \sum_i \lambda_i$. Let $\vec{\delta}$ be the neighbor vectors, then the mean field Hamiltonian is given by 
\begin{eqnarray}
H_{\text{MF}}  & = & \sum_{i ,\delta}\Big[ \cos\theta  \hat{B}^{\dagger}_{i, i+\delta} \mathbf{B}^\prime(\delta) + \sin\theta  \hat{C}^{\dagger}_{i, i+\delta} \mathbf{B}^\prime(\delta)   \nonumber \\
& - & \cos\theta  \hat{A}^{\dagger}_{i, i+\delta} \mathbf{A}^\prime(\delta) +  \sin\theta  \hat{D}^{\dagger}_{i, i+\delta} \mathbf{A}^\prime(\delta)]
 +  \text{H.C.} \Big] + \epsilon_0\nonumber
\end{eqnarray}
 We have, define the Fourier transformation $ \mathbf{\Phi}_i = \frac{1}{\sqrt{N_u}} \sum_\mathbf{k} e^{-i \mathbf{k} \cdot \mathbf{r}_i} \mathbf{\xi}_{\mathbf{k},\mu_i}$ where $\mathbf{\xi}_{\mathbf{k},\mu_i} = \begin{pmatrix} \alpha_{\mathbf{k}, \mu_i} \\ \beta_{\mathbf{k}, \mu_i} \end{pmatrix} $ with  sub-lattice index $\mu_i$ and $N_u$ be the total number of unit cells. Then  above mean field Hamiltonian reduces to a compact form, given by
\begin{equation}
H_{\text{MF}} = \sum_{\mathbf{k} > 0} \Psi^\dagger_\mathbf{k} D_\mathbf{k} \Psi_\mathbf{k} + \epsilon_0
\end{equation}
We define $ \Psi^T_{\mathbf{k}}=\begin{pmatrix}\mathbf{\xi}_{\mathbf{k},1}, \mathbf{\xi}_{-\mathbf{k},1}^{\dagger}, \mathbf{\xi}_{\mathbf{k},2},\mathbf{\xi}_{-\mathbf{k},2}^{\dagger},\mathbf{\xi}_{\mathbf{k},3},\mathbf{\xi}_{-\mathbf{k},3}^{\dagger}\end{pmatrix} $ and the $D_\mathbf{k}$ matrix is given by
\begin{widetext}
\begin{equation} 
D_\mathbf{k}=\begin{pmatrix}
-\lambda & 0 & Y_{12}(\mathbf{k}) & X_{12}(\mathbf{k}) & Y^{\dagger}_{31}(\mathbf{k}) & X^{T}_{31}(-\mathbf{k})\\
 	0 & -\lambda & X_{12}(-\mathbf{k})^{\dagger T} & Y_{12}(-\mathbf{k})^{T\dagger} & X_{31}(\mathbf{k})^{\dagger} & Y_{31}(-\mathbf{k})^{T}\\
 	Y_{12}(\mathbf{k})^{\dagger} & X_{12}(-\mathbf{k})^{T} & -\lambda & 0 & Y_{23}(\mathbf{k}) & X_{23}(\mathbf{k})\\
 	X_{12}(\mathbf{k})^{\dagger} & Y_{12}(-\mathbf{k})^{T} & 0 & -\lambda & X_{23}(-\mathbf{k})^{\dagger T} & Y_{23}(-\mathbf{k})^{T\dagger}\\
 	Y_{31}(\mathbf{k}) & X_{31}(\mathbf{k}) & Y_{23}(\mathbf{k})^{\dagger} & X_{23}(-\mathbf{k})^{T} & -\lambda & 0\\
	X_{31}(-\mathbf{k})^{\dagger T} & Y_{31}(-\mathbf{k})^{T\dagger} & X_{23}(\mathbf{k})^{\dagger} & Y_{23}(-\mathbf{k})^{T} & 0 & -\lambda
 \end{pmatrix}
 \end{equation}
 where $X_{ij}(\mathbf{k})$ and $Y_{ij}(\mathbf{k})$ are are $2 \times 2$ matrix, given by
 \begin{eqnarray}
X_{ij}(\mathbf{k})&=\frac{i}{2}\left(\mathbf{A}_{ij}^\prime e^{-i\mathbf{k}\cdot \mathbf{r}_{ij}}+\mathbf{A}_{i+3,j+3}^\prime e^{-i\mathbf{k}\cdot \mathbf{r}_{i+3,j+3}}\right)\sigma_{y}\left(\cos\theta-i\sin\theta~\hat{\mathbf{d}_{ij}}\cdot\sigma\right)\\
Y_{ij}(\mathbf{k})&=\frac{1}{2}\left(\mathbf{B}_{ij}^\prime e^{-i\mathbf{k}\cdot \mathbf{r}_{ij}}+\mathbf{B}_{i+3,j+3}^\prime e^{-i\mathbf{k}\cdot \mathbf{r}_{i+3,j+3}}\right)\left(\cos\theta+i\sin\theta~\hat{\mathbf{d}_{ij}}\cdot\sigma\right)
 \end{eqnarray}
  \end{widetext}
The structure of the $D_\mathbf{k}$ matrix is slightly different at the special points $\Gamma, M$ and $K$,where $\Gamma = (0,0), M = (0, 2 \pi/\sqrt{3})$, $K = (1,\sqrt{3})2 \pi/3$, $ M_e= (0, 4 \pi/\sqrt{3})$ and $ K_e = (1,\sqrt{3})4 \pi/3$.   Using the standard Bigoliouubov transfromation we can diagonalize the mean field Hamiltonian  $\Psi_\mathbf{k} = M \tilde{\Psi}_\mathbf{k}$ where $M $ is the Bigoliouubov matrix  of the  form
$ M= \begin{pmatrix} 	U  & V \\	X  & Y \end{pmatrix}$
 and $ \tilde{\Psi}^T_{k}=\begin{pmatrix}\tilde{\mathbf{\xi}}_{\mathbf{k},1},\tilde{\mathbf{\xi}}_{-\mathbf{k},1}^{\dagger}, \tilde{\mathbf{\xi}}_{\mathbf{k},2},\tilde{\mathbf{\xi}}_{-\mathbf{k},2}^{\dagger},\tilde{\mathbf{\xi}}_{\mathbf{k},3},\tilde{\mathbf{\xi}}_{-\mathbf{k},3}^{\dagger}\end{pmatrix} $ and  $\tilde{\mathbf{\xi}}_\mathbf{k} = \begin{pmatrix} \tilde{\alpha}_\mathbf{k} \\ \tilde{\beta}_\mathbf{k} \end{pmatrix}$
 
 \vspace{0.5cm}
The mean field energy is given by
\begin{eqnarray}
E_{\text{MF}}  & = & \sum_{\mu ,\mathbf{k} >0}\Big [\omega_{\mathbf{k} \mu} (\tilde{\alpha}^\dagger_{\mathbf{k} \mu} \tilde{\alpha}_{\mathbf{k} \mu} + \tilde{\beta}^\dagger_{\mathbf{k} \mu} \tilde{\beta}_{\mathbf{k} \mu}) +(2 S +1) N \lambda \nonumber \\
& + & 2 (\mathbf{A}^{\prime 2} - \mathbf{B}^{\prime 2})\Big]
\end{eqnarray}
where $\omega_{\mathbf{k} \mu}$ is the dispersion relation of the $\mu =1,.......,2m$ spinon modes with $m$ be the number sites within the unit cell. The ground state $|\tilde{0}\rangle$ is the vacuum of the Bogoliubov bosons. 
\vspace{0.25cm}

The mean field parameters can be obtained by extrimizing the mean field energy with respect to the mean field parameters which is equivalent to the solve the self-consistency equations.
\begin{equation}
\frac{\partial E}{ \partial \mathbf{A}^\prime} = 0 ,\hspace{0.5cm} \frac{\partial E}{ \partial \mathbf{B}^\prime} = 0 \hspace{0.5cm}\text{and} \hspace{0.5cm} \frac{\partial E}{ \partial \lambda} = 0
\end{equation}
The optimisation process is computationally difficult due to the skewed nature the Hessian at the saddle point~\cite{PhysRevB.95.134404} near the phase transitions. We have used the classical ground state to estimate the initial guesses for the mean fields.

\begin{figure*}[ht]
\begin{center}
\includegraphics[height=5.5cm,width=5.9cm]{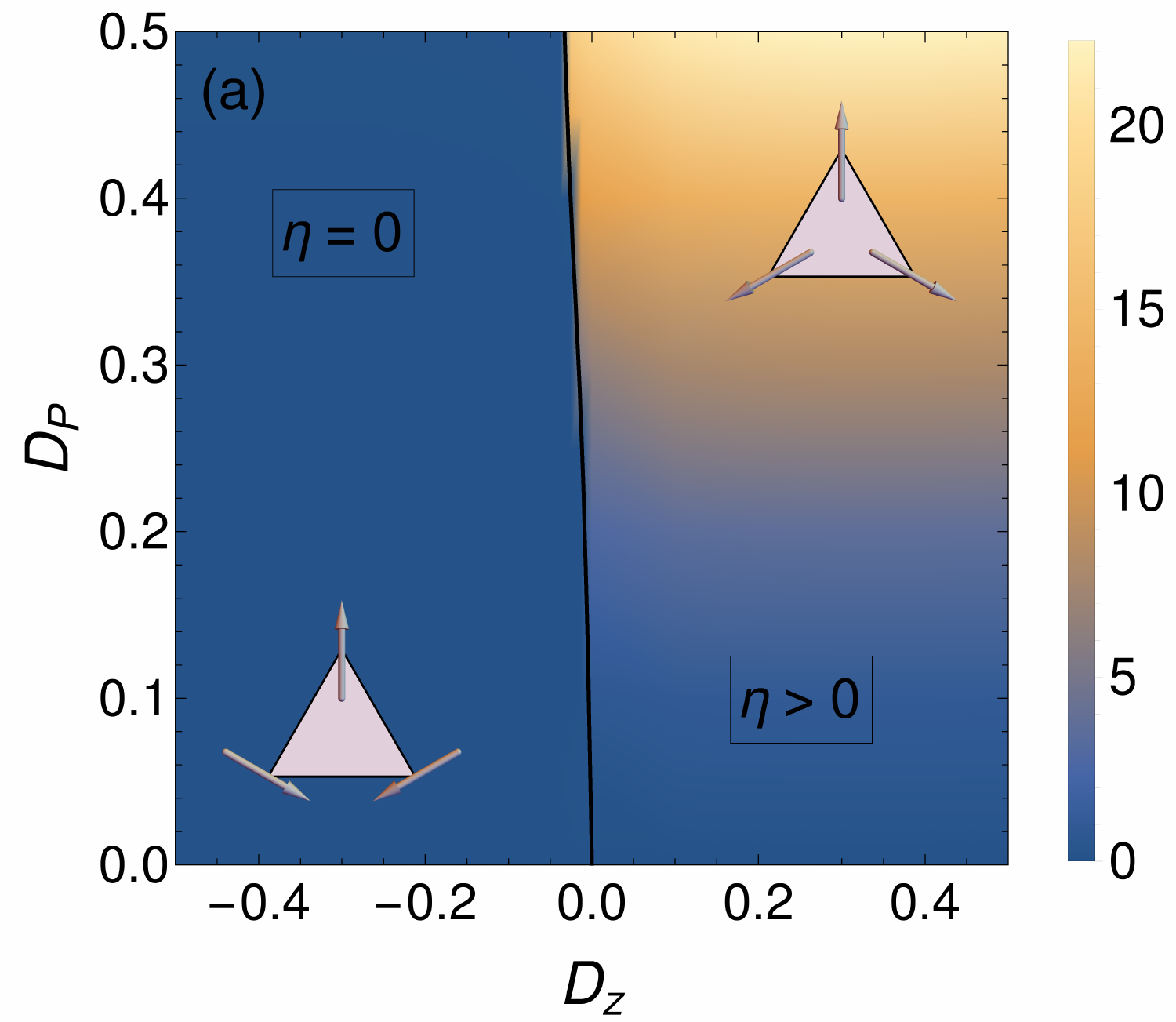}
\includegraphics[height=5.5cm,width=5.9cm]{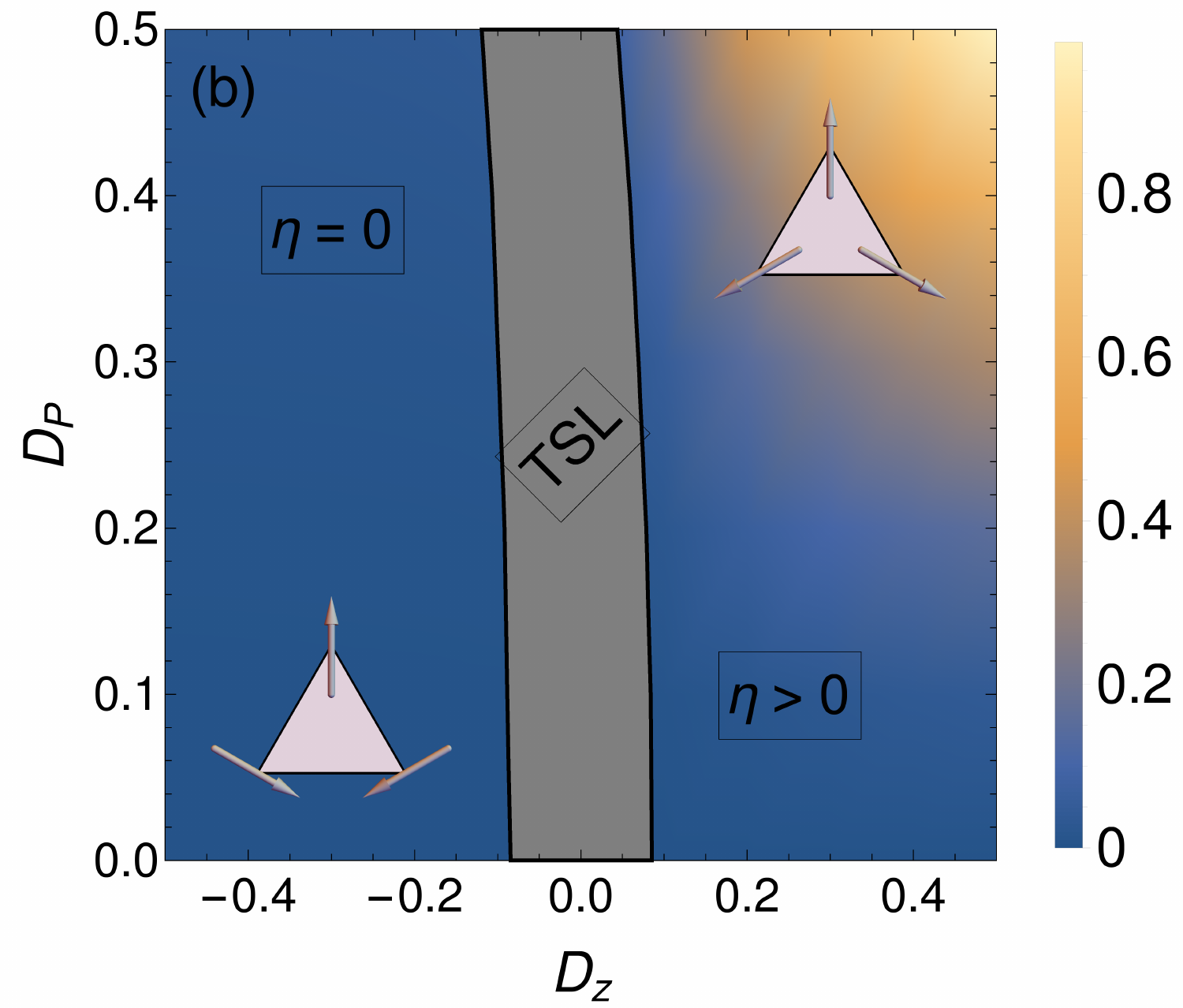} 
\includegraphics[height=5.5cm,width=5.9cm]{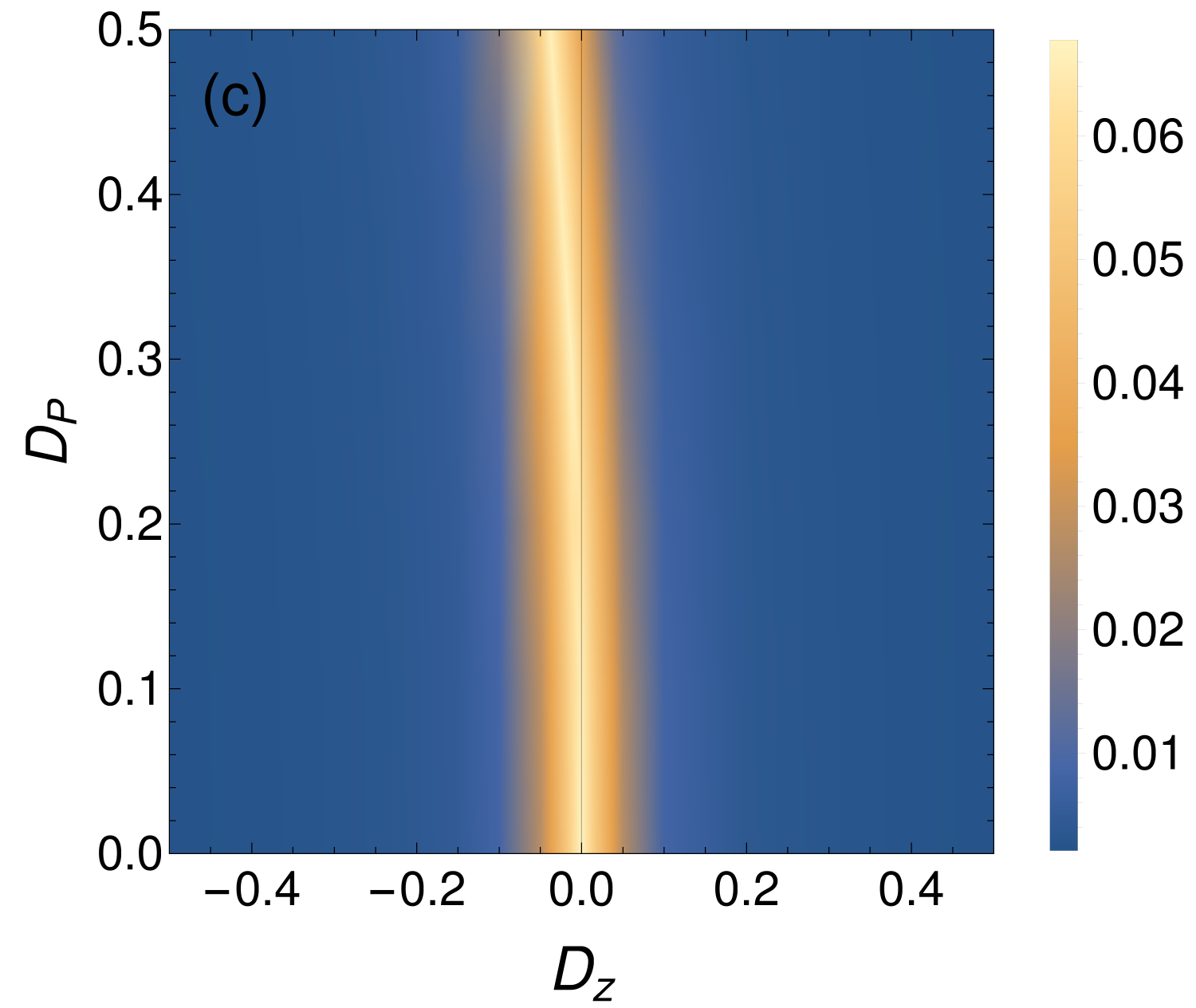}
\caption{Ground state phase diagram for (a) $S = 0.5$ and  (b) $S =0.2$ (c) Gap as a function of $D_p$ and $D_z$ for $S =0.2$.}
\label{fig: sbmft-pd-gap}
\end{center}
\end{figure*}

\begin{figure*}[ht]
\includegraphics[height=5.2cm,width=5cm]{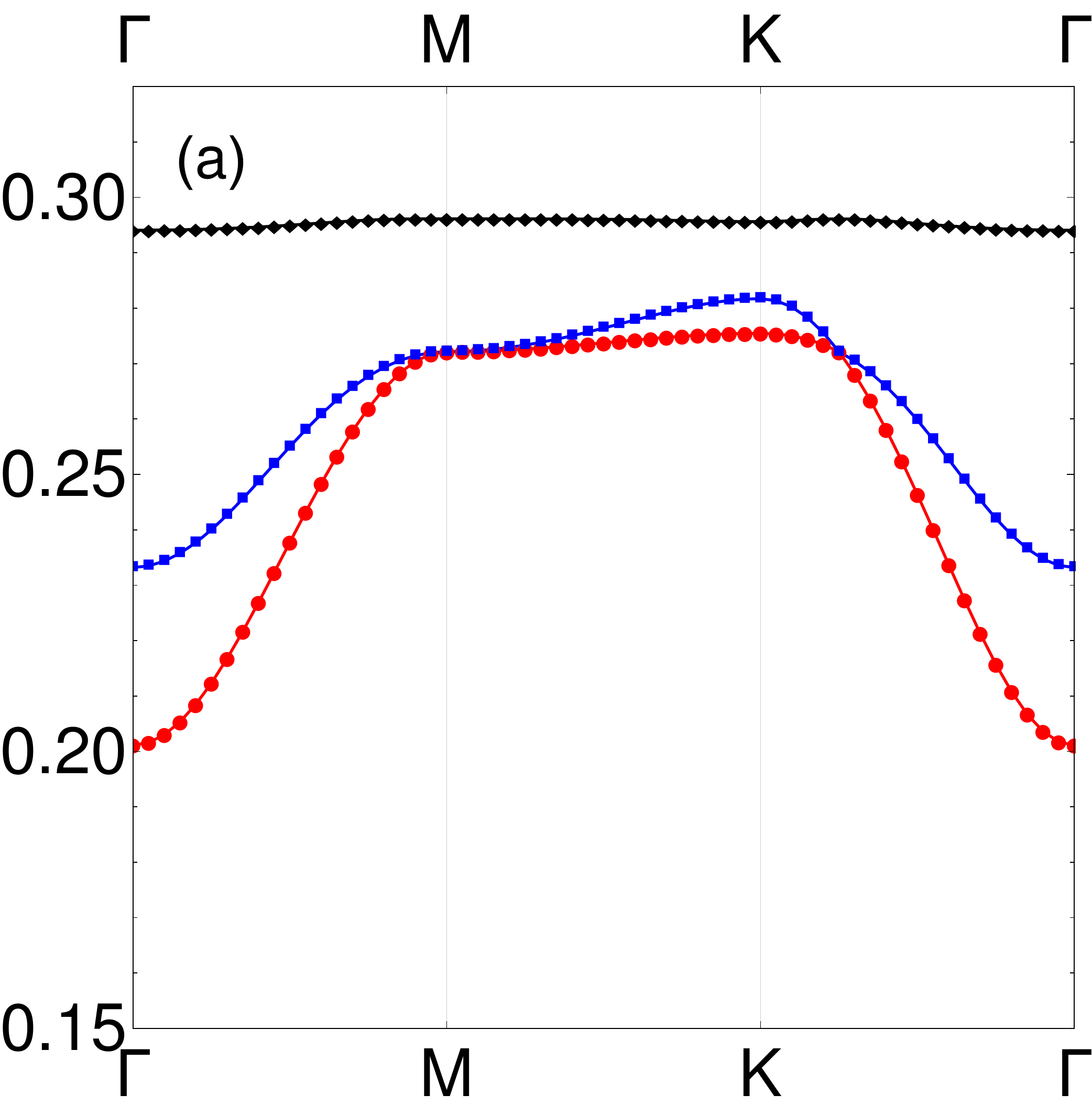}\hspace{1cm}
\includegraphics[height=5.2cm,width=5cm]{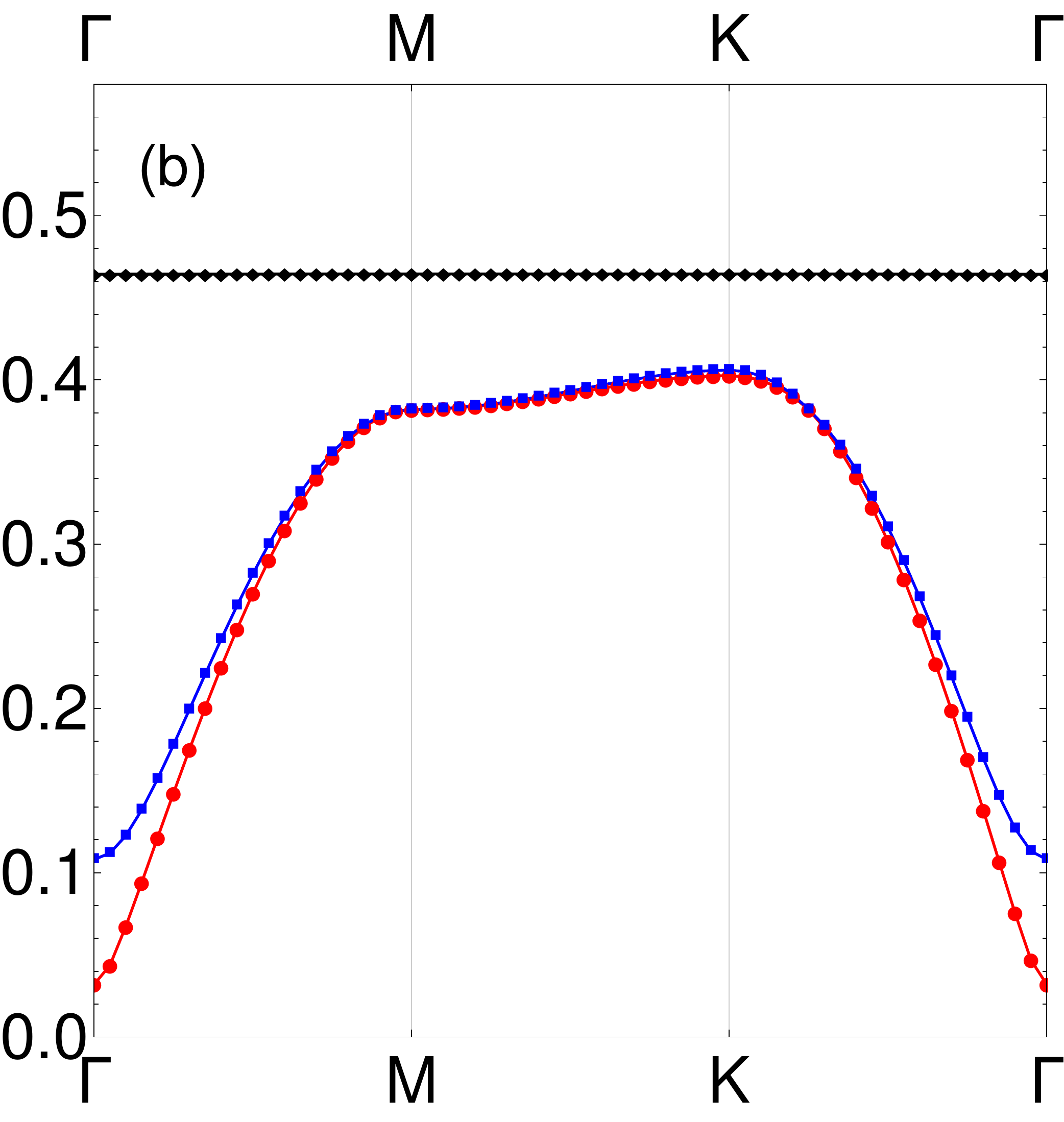}\hspace{1cm}
\includegraphics[height=5.2cm,width=5cm]{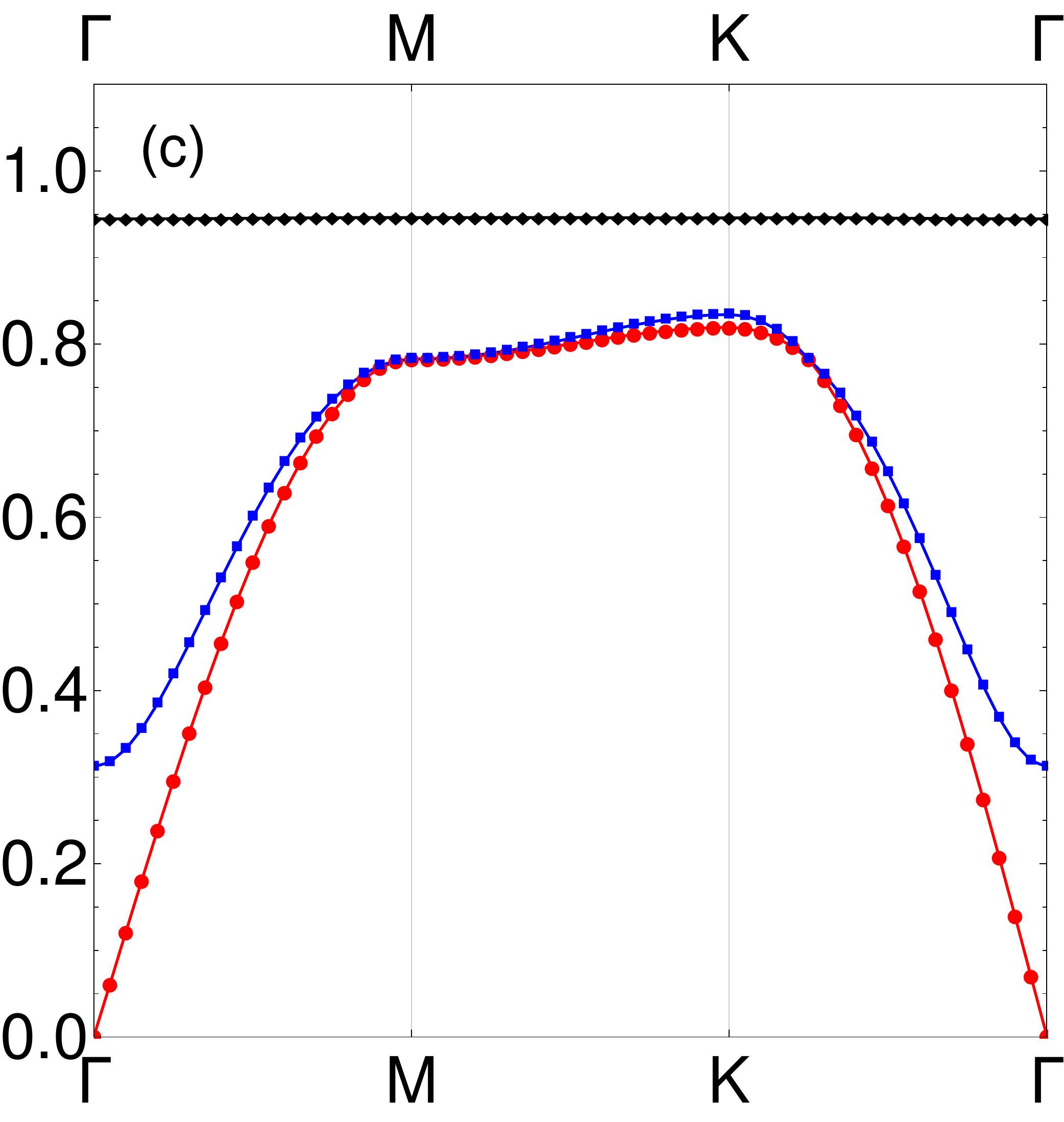}
\caption{ Spinon spectrum in (a) spin liquid region at $S = 0.05$ with $D_p = 0.05$ and $D_z  =- 0.3$ and (b) spin liquid region at $S=0.2$ with $D_p = 0.2$ and $D_z =-0.05$ (c) LRO region $S = 0.5$ with $D_p = 0.05$ and $D_z  = 0.1$  }
\label{fig:sbmft-spectrum}
\end{figure*}

\vspace{0.5cm}
\noindent \textbf{Initial guess for mean fields :} In the large $S$ limit the SBMFT result should mimic the classical ground state i.e the $\mathbf{Q=0}$ umbrella state. In the classical limit we define

 \begin{subequations}
\begin{eqnarray}
\langle a \rangle & = & r_1 e^{i \nu_1}  \\
\langle b \rangle & = & r_2 e^{i \nu_2} 
\end{eqnarray}
 \end{subequations}
 
where $r_1, r_2$ is the modulus and $\nu_1, \nu_2$ be the argument  of the average values of two flavors of bosonic operator  $a $ and $b$  in the classical limit. To obtain the $\mathbf{Q=0}$  spin configuration, we must have

 \begin{subequations}
\begin{eqnarray} 
\left\langle a_{i}\right\rangle 	& = & \sqrt{2S}\cos\left(\frac{\zeta}{2}\right)  \\
\left\langle b_{i}\right\rangle & = & \sqrt{2S}\sin\left(\frac{\zeta}{2}\right)e^{i\beta_{i}}
\end{eqnarray}
 \end{subequations}
 
where $ \beta_{1}=\frac{7\pi}{6}, \beta_{2}=\frac{\pi}{2}$ , and  $\beta_{3}=-\frac{\pi}{6}$ and $\zeta = \pi/2 - \eta$. The mean fields we obtain using these are summarized in Table.~\ref{tbl:bond-operator} \\
\begin{table}[ht]
\begin{center}
\begin{tabular}{|c|c|c|c|c|c|}
 \hline
 \multicolumn{2}{|c|}{Mean-fields}           &  Bond-(1,2)   &  Bond-(2,3)  &   Bond-(3,1)   &  After Gauge  \\ 
 \hline
\multirow{2}*{~~A~~}  & $\left|\cdot\right|$  &        \multicolumn{3}{c|}{$\frac{S}{2}\sqrt{3}\sin(\zeta)$}  & -\\  \cline{2-6}
                               &         ~Phase~                    &    $\frac{\pi}{3}$ & $-\frac{\pi}{3}$ & $\pi$         & 0 \\
\hline    
\multirow{2}*{~~D~~}  & $\left|\cdot\right|$  &        \multicolumn{3}{c|}{$\frac{S}{2}\left(2D_{p}\cos(\zeta)+D_{z}\sin(\zeta)\right)$}  &  - \\  \cline{2-6}
                               &           ~Phase~                    &  $\pi+\frac{\pi}{3}$ & $\pi-\frac{\pi}{3}$ & $0$           & $\pi$  \\
\hline  
\multirow{2}*{~~B~~}  & $\left|\cdot\right|$  &          \multicolumn{3}{c|}{$\frac{S}{2}\sqrt{3\cos^{2}(\zeta)+1}$}& - \\  \cline{2-6}
                               &           ~Phase~                    &             $\Phi_{B}$          &           $\Phi_{B}$       &          $\Phi_{B}$           & $\Phi_{B}+\frac{4\pi}{3}$ \\
\hline    
  \multirow{2}*{~~C~~}  & $\left|\cdot\right|$  &         \multicolumn{3}{c|}{$\frac{S}{2}\sqrt{\left(D_{z}\cos(\zeta)-2D_{p}\sin(\zeta)\right)^{2}+3D_{z}^{2}}$}&-\\  \cline{2-6}
                               &           ~Phase~                    &             $\Phi_{C}$         &           $\Phi_{C}$           &          $\Phi_{C}$             &$\Phi_{C}+\frac{4\pi}{3}$  \\
\hline    
\end{tabular}
\caption{
Phases and magnitudes (denoted by $|.|$) of  different mean fields, defined in
 Eq.~\ref {eqn:definitionAandB} and Eq.~\ref{eqn:definitionCandD}  for bonds $(1,2),\,(2,3)$ and $(3,1)$ shown in Fig.~\ref{DM-config-classical-pd}(a) in the classical limit.}
\label{tbl:bond-operator}
\end{center}
\end{table}

\vspace{-0.5cm}
Now, we use the gauge transformation  $a_i \rightarrow e^{i \phi(i)} a_i$ and $b_i \rightarrow e^{i \phi(i)} b_i$ with the phases $\phi(i)$ set to $\frac{5\pi}{6},\, -\frac{\pi}{2},\,\frac{\pi}{6} $ at three sublattices respectively to transform both $A$ and $D$ into real fields. At the same time, $B$ and $C$ fields will acquire a constant phase of $4\pi/3$. Thus, in the final calculation, we can take $\mathbf{ A}^\prime$ as a real number and $\mathbf{ B}^\prime$ as a complex number. However, the optimization of the mean-field parameters shows that for the symmetry of the spiral order, we must have $\mathbf{ B}^\prime=0$. We are left to optimize $\mathbf{ A}^\prime$ field and $\lambda$.

\vspace{0.25cm}
With these as our initial guess, we optimize the mean field parameters. A few sample values of mean fields and energies are given in  Table~\ref{table:nonlin} for $N=1200$ (20 unitcells $\times$ 20 unitcells).
\begin{table}[ht]
\centering
\begin{tabular}{|c| c| c|c| c| c|}
\hline
\textit{S} & $D_p$ & $D_z$ & $\mathbf{A}^\prime$ &$\lambda$ & Energy\\ [0.5ex]
\hline
0.20& 0.2 &0.05& 0.26429&-0.46458&-0.13970\\
\hline
0.50 & 0.2 & 0.05& 0.52736&-0.92482& -0.55622\\ [1ex] 
\hline
\end{tabular}
\caption{Optimized values of the mean field parameters and energies for  $N=1200$}
\label{table:nonlin}
\end{table}
\section{SBMFT results}
\begin{figure}[ht!]
\includegraphics[height=4.75cm,width=6.75cm]{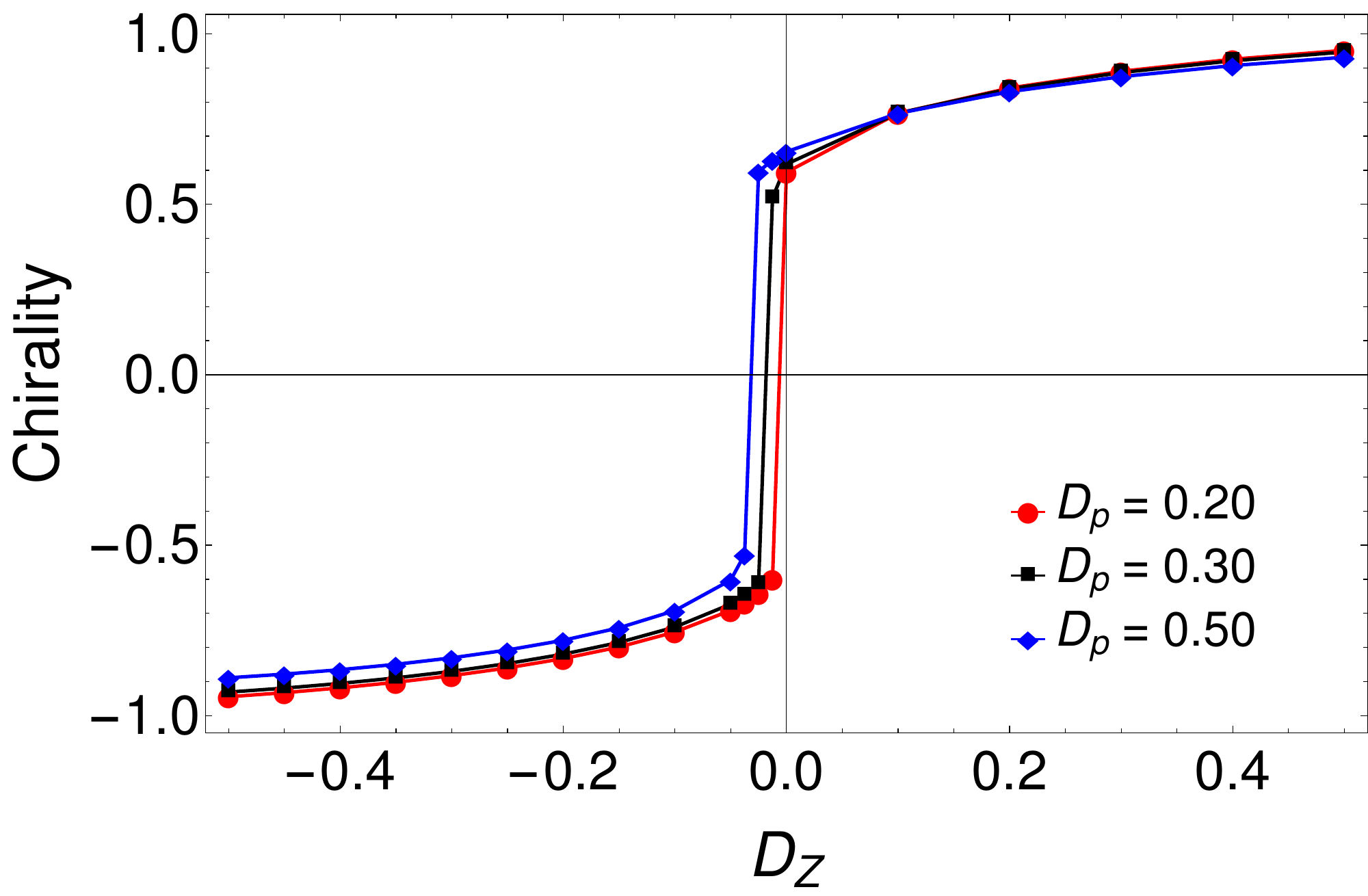}
\caption{ Chirality as a function of $D_z$ for different values of $D_p$ at $S = 0.5$}
\label{fig:sbmft-Chirality}
\end{figure}

We have computed the zero-temperature ground state phase diagram for this model in the parameter space of $0 \leq D_p \leq 0.5$ and $-0.5 \leq D_z \leq 0.5$ based on chirality, the ZZ correlation at $\Gamma$ point and also the gap in the thermodynamic limit for various values of $S$. Our proposed ground state phase diagram for $S =0.5$ is quite similar to the classical phase diagram and is, as shown in Fig.~\ref{fig: sbmft-pd-gap}(a). In the phase diagram for $S = 0.5$, we found $\mathbf{Q=0}$ structure with two chiralities. The boundary between two phases obtained from the chirality is shown by a black line. The boundary obtained from SBMFT is not exactly the same as obtained from the classical case, but the qualitative features remain the same as shown in Fig.~\ref{fig: sbmft-pd-gap}(a). The boundary line is curved into the first quadrant of the phase diagram; that is, the chirality changes at negative values of $D_z$ for a given larger $D_p$, as shown in Fig.~\ref{fig:sbmft-Chirality}. This is a result of the fact that the chirality selected by these two components is not the same, and hence there is a competition between these two. In the negative chirality phase, the spins are forced to lie in the kagome plane, resulting in $S^z =0$, whereas for the positive chirality, the spins are canted and the canting angle is shown by the color gradient. If we change the sign the in-plane component, the effect of $D_p$ is to change the canting angle from positive to negative.

\begin{figure*}[t]
\includegraphics[height=5.45cm,width=5.85cm]{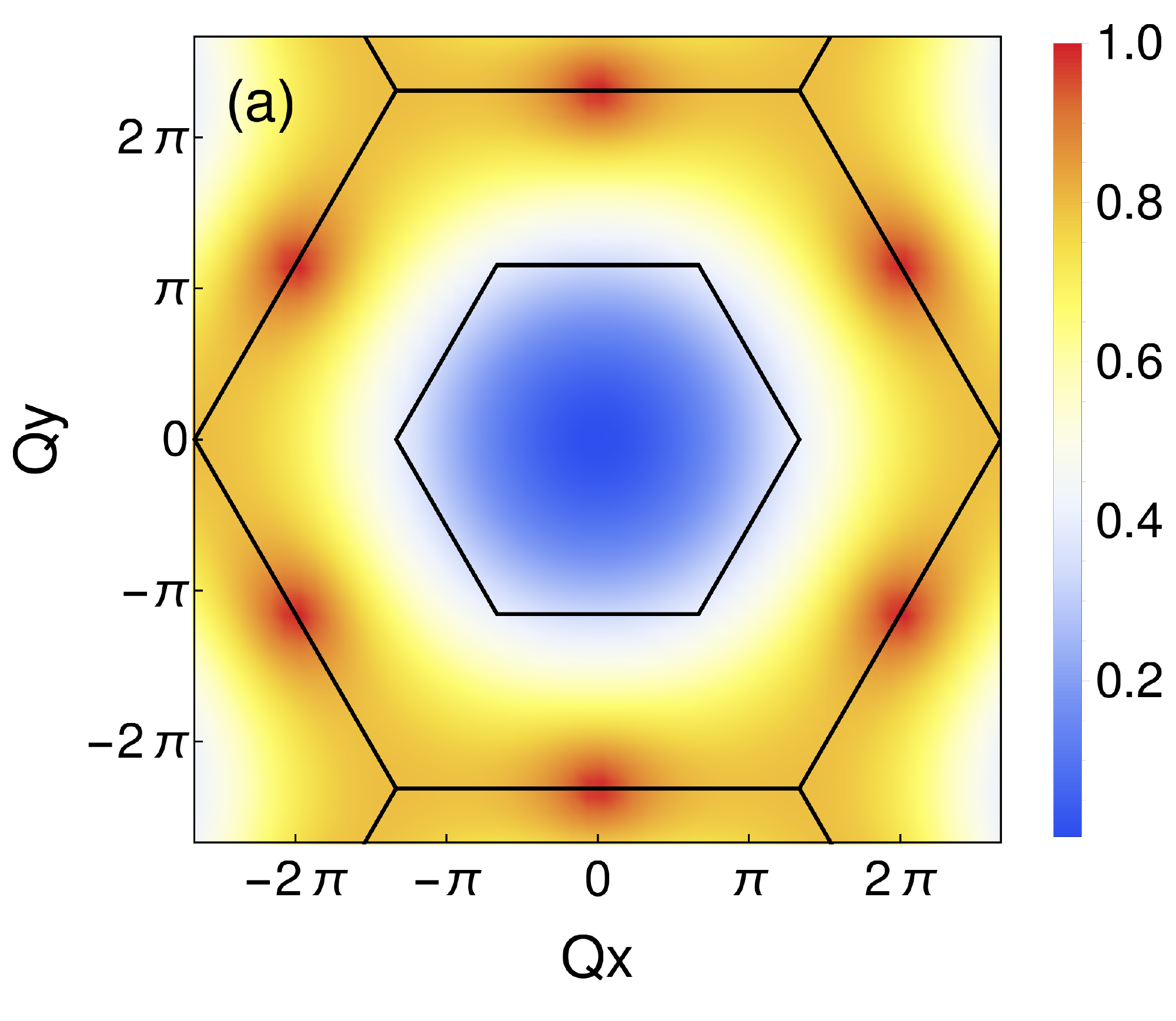}\hspace{0.075cm}
\includegraphics[height=5.5cm,width=5.85cm]{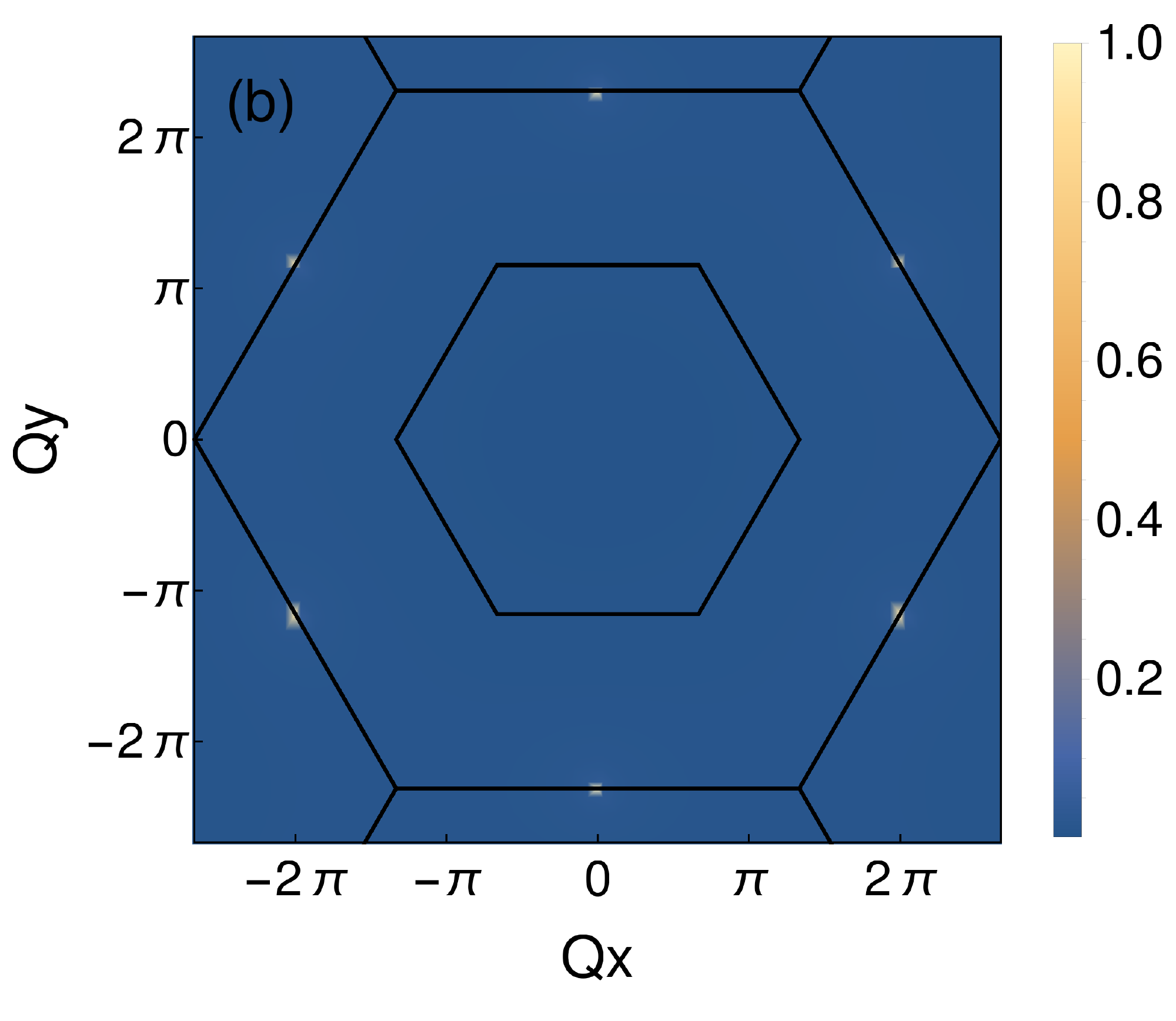}\hspace{0.075cm}
\includegraphics[height=5.5cm,width=5.85cm]{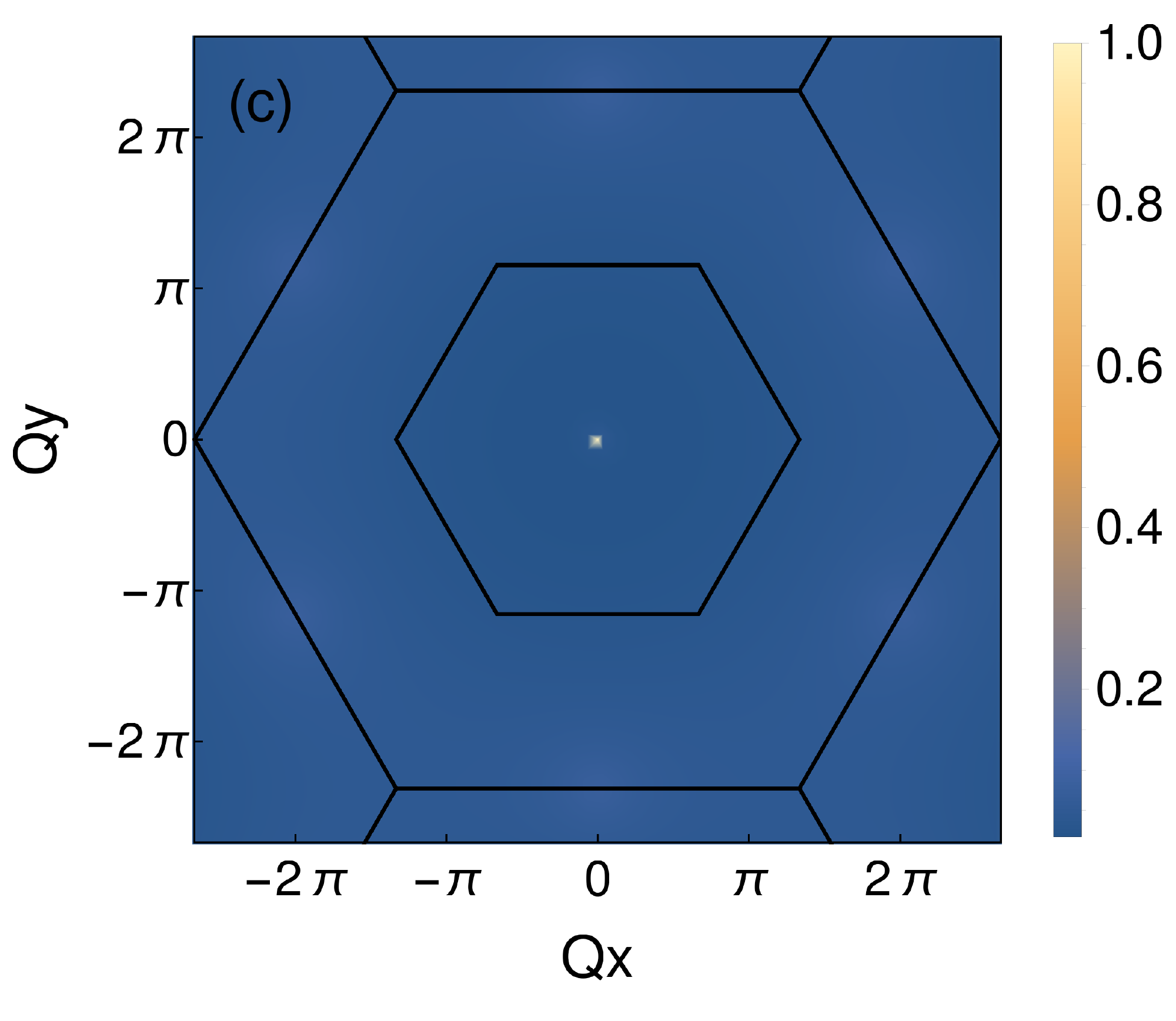}
\caption{(a) XX-component of SSF at $S = 0.2$ with $D_p =0.05$ and $D_z =- 0.05$ (b) XX-component of SSF at $S = 0.5$ with $D_p =0.5$ and $D_z = 0$ (c) ZZ-component of SSF at $S = 0.5$ with $D_p =0.5$ and $D_z = 0$ }
\label{fig:sbmft_SSF}
\end{figure*}

As the value of $S$ is lowered, a gapped spin liquid phase opens up at the boundary line at a critical value of $S$, which is very close to $0.2$, and the spin-liquid phase becomes wider with decreasing value of $S$. The ground-state phase diagram for $S= 0.2$ is shown in the Fig.~\ref{fig: sbmft-pd-gap}(b) At this value of $S$, the gapped spin liquid phase is sandwiched between two $\mathbf{Q=0}$ LRO phases with two different chirality. The boundary between the QSL and the LRO is obtained from the extrapolation of the gap data. Gap as a function of $D_p$ and $D_z$ is shown in the density plot in Fig.~\ref{fig: sbmft-pd-gap}(c)

\vspace{0.25cm}

Fig.~\ref{fig:sbmft-spectrum} shows the quasi particle dispersion relations along the high symmetry line $\Gamma$ - M - K -$ \Gamma$.  The topological spin-liquid(TSL) phase is characterized by a gapped spinon spectrum, whereas in the magnetic long-range order state, the spectrum is gapless at the thermodynamic limit. The low energy excitations in long-range order states are magnons, which can be viewed as the bound state of two spinons, glued together~\cite{sachdev1992kagome,PhysRevB.98.184403}.   In Fig.~\ref{fig:sbmft-spectrum}(a) the spinon spectrum is shown at $S=0.05$ with $D_p=0.05$ and $D_z =-0.3$. The spectrum is gapped indicating the ground state is in the spin liquid state. The spectrum for $S=0.2$ with $D_p = 0.2$ and $D_z = -0.05$ is as shown in the Fig.~\ref{fig:sbmft-spectrum}(b) where the spectrum is still gapped. In Fig.~\ref{fig:sbmft-spectrum}(c) the spinon spectrum is shown for $S=0.5$ with $D_p = 0.05$ and $D_z=0.1$ which is gapless at $\Gamma$ indicating  the ground state has acquired LRO. 

To investigate the long range magnetic order, we calculate the static structure factor which is defined as 
\begin{equation}
S^{\alpha \beta}(\mathbf{Q}) = \frac{3}{4 N}\sum_{ij} e^{i \mathbf{Q}\cdot (\mathbf{R}_i - \mathbf{R}_j)} \langle 0 | S^\alpha_i S^\beta_j | 0 \rangle
\label{eqn:sf}
\end{equation}
where $R_i$ and $R_j$ is the site index and $\alpha, \beta \in \{x,y,z\}$. A magnetic long order state produces a sharp discrete Bragg peaks where as QSL produces a continuous, diffusive scattering spectra. Here, we have calculated both the transverse component and ZZ component of static structure factor for different points in the parameter space to examine the magnetic structure.

The XX-component of static structure factor for $S = 0.20$ at $D_p =0.05$ and $D_z =-0.05$ is shown in the Fig.~\ref{fig:sbmft_SSF}(a). This is a representative point in the QSL region of the phase diagram. There is a broad peak at $M_e$ supports the conclusion that the ground state is in the spin-liquid state. To illustrate the canting of the spins, we show the static structure factor for $S =0.5$ with $D_z = 0$ and $D_p =0.5$ in Fig.~\ref{fig:sbmft_SSF}(b). Sharp peaks appear at the $M_e$ point, indicating a magnetic LRO of $\mathbf{Q=0}$ type. From Fig.~\ref{fig:sbmft_SSF}(c), we see a peak at $\Gamma$ point. This suggests that there is a ferromagnetic component along the z-direction, which is the result of the spins tilting away from XY plane. The canting angle can also be estimated from this data.
\begin{figure*}[ht!]
\begin{center}
\includegraphics[height=6.5cm,width=7.5cm]{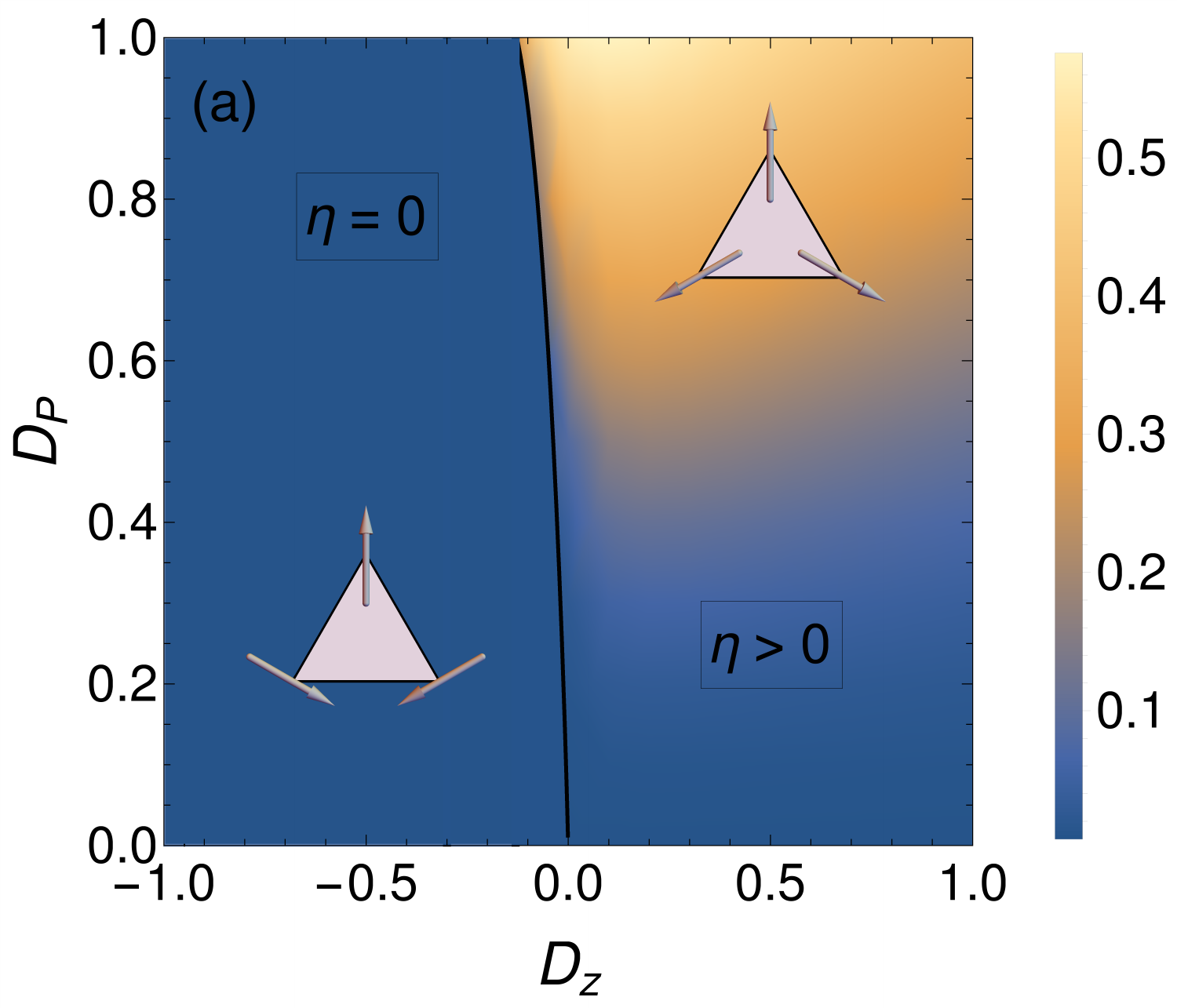}\hspace{0.5cm}
\includegraphics[height=6cm,width=8.5cm]{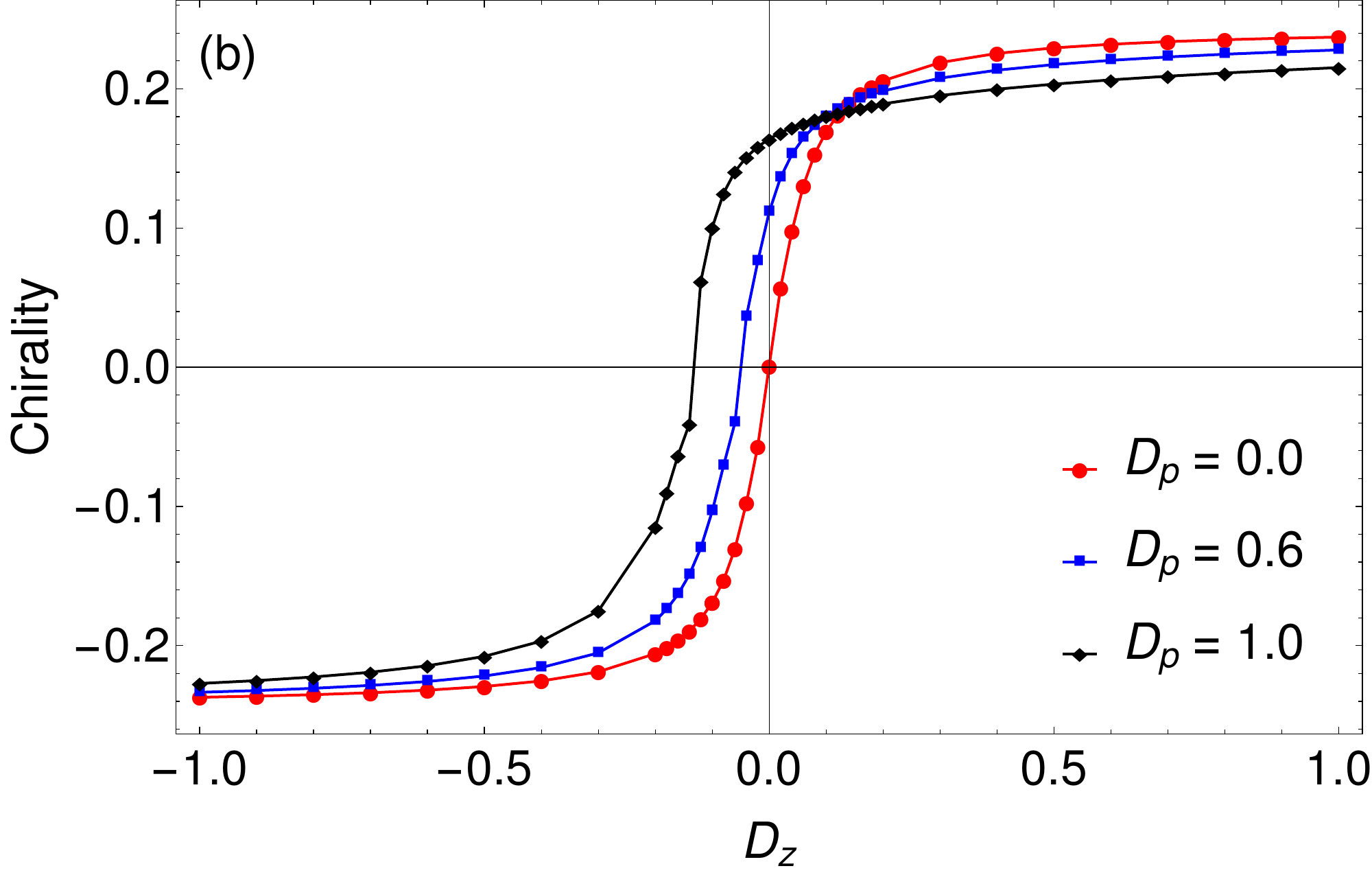}
\end{center}
\caption{ (a) Ground state phase diagram and (b) Chirality as a function of $D_z$ for different values of $ D_p$  for N=27}
\label{fig:PD_ED-chirality-ED}
\end{figure*}
\section{Exact diagonalization study} 
To verify the results obtained using the SBMFT, we have computed numerical results by employing the
exact diagonalization method up to 30 sites.
We have computed various physical quantities to examine the magnetic structure of this model.
This method is exact and widely used to study the ground state of different frustrated magnets though it is limited by small system size due to the huge computational requirement. In addition, the absence of the global spin rotation symmetry, due to the presence of DM interaction, the Hilbert space could not be decomposed into the invariant subspaces, restricting the size of the system to 30.
In our computation, We have used the package PARPACK to diagonalize the sparse Hamiltonian matrix with a total number of spins $N$ is $12,15,18,21,24,27,30$ for different shapes. We have applied periodic boundary conditions to reduce the finite size effect. We have obtained the ground-state phase diagram of the model only for the positive $D_p$. We must mention that the conclusion is made by looking at the trends of data points and extrapolation, so the accurate location of the critical point largely depends on the extrapolation function as well as cluster size and shape.
\vspace{0.25cm}

\begin{figure}[h]
\includegraphics[height=6cm,width=8cm]{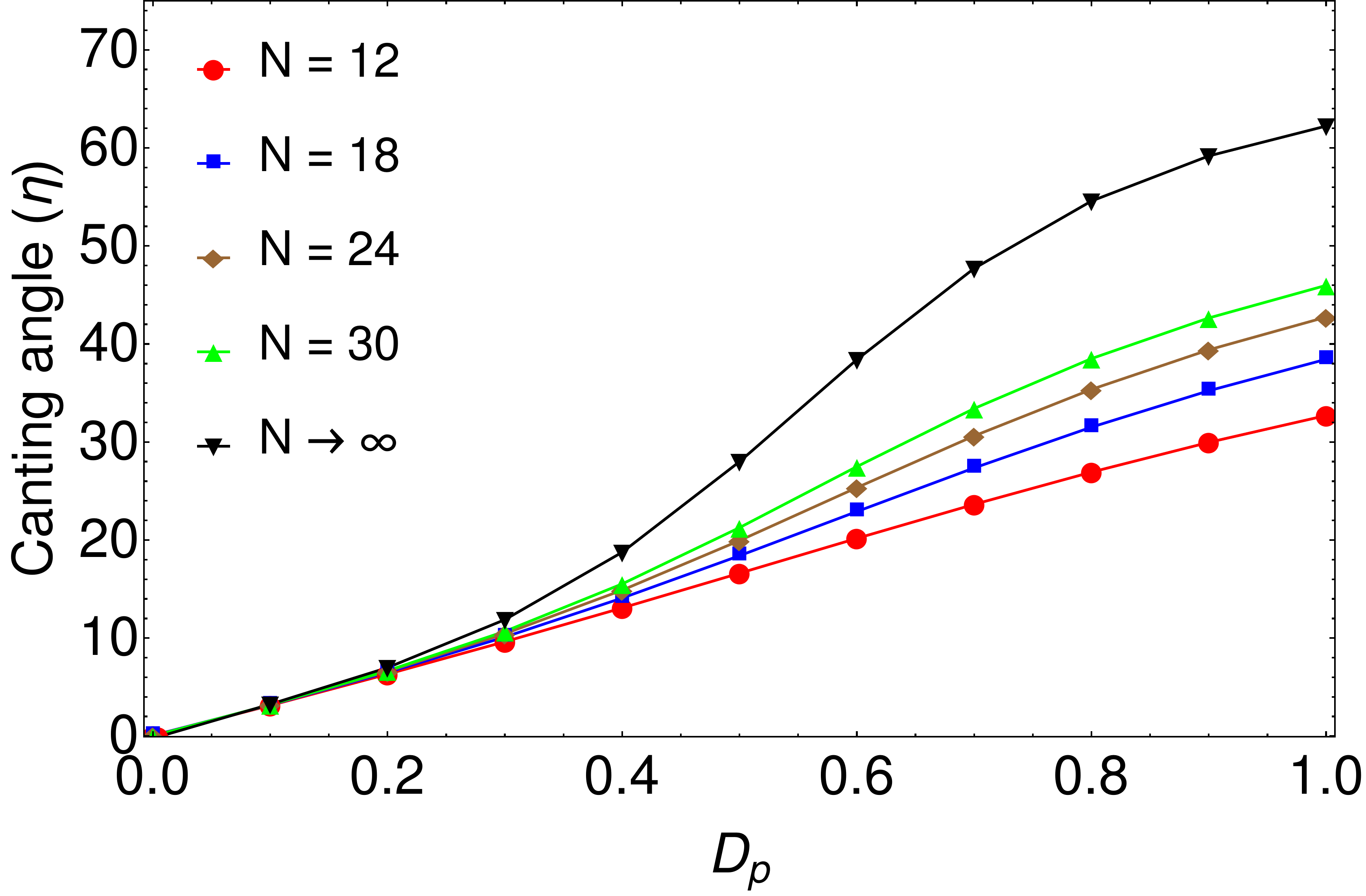}
\caption{Canting angle(in degree) as a function of $D_p$ for $D_z=0.1$}
\label{fig:ca}
\end{figure}

The ground-state phase diagram of the present model is presented in Fig.~\ref{fig:PD_ED-chirality-ED}(a). The ground state is in the long-range order throughout the parameter space. For $D_p = 0$ case is discussed by Cepas et al.~\citep{cepas2008quantum} and they give evidence for spin liquid below $D_z = 0.1J$ using the idea of tower of states. However, as soon as $D_p$ is added, the global $U(1)$ symmetry vanishes, and hence the same idea cannot be used to determine the existence of spin liquid. Away from the isotropic point, the ground state is in $\mathbf{Q=0}$ LRO. The ground state is planar with negative chirality when $D_z < 0 $. With $D_z > 0$, the state is an umbrella state with positive chirality. In this case, the canting angle varies from zero to $62.23^o$ as $D_p$ changes from zero to $J$, as shown in Fig.~\ref{fig:ca}. These results are in agreement with the SBMFT results presented earlier.

The Fig.~\ref{fig:PD_ED-chirality-ED}(b) shows the chirality as a function of $D_z$ for different values of $D_p$ for $N=27$. The crossover of chirality shifts to positive $D_z$ as $D_p$ increases in value. This crossover point is used to distinguish between the phases in the phase diagram. 
\begin{figure*}[ht!]
\begin{center}
\includegraphics[height=5.5cm,width=8cm]{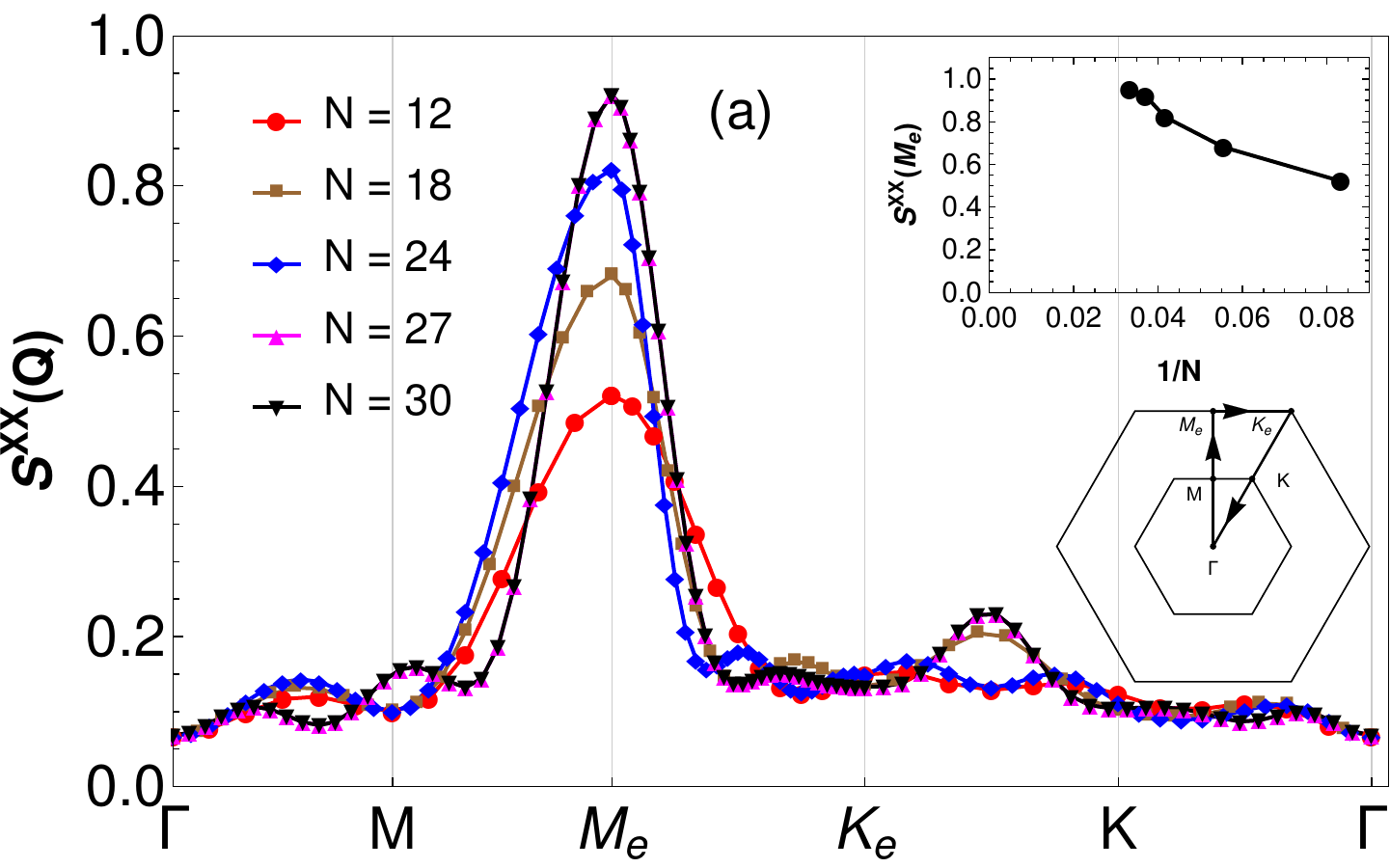}\hspace{1cm}
\includegraphics[height=5.5cm,width=8cm]{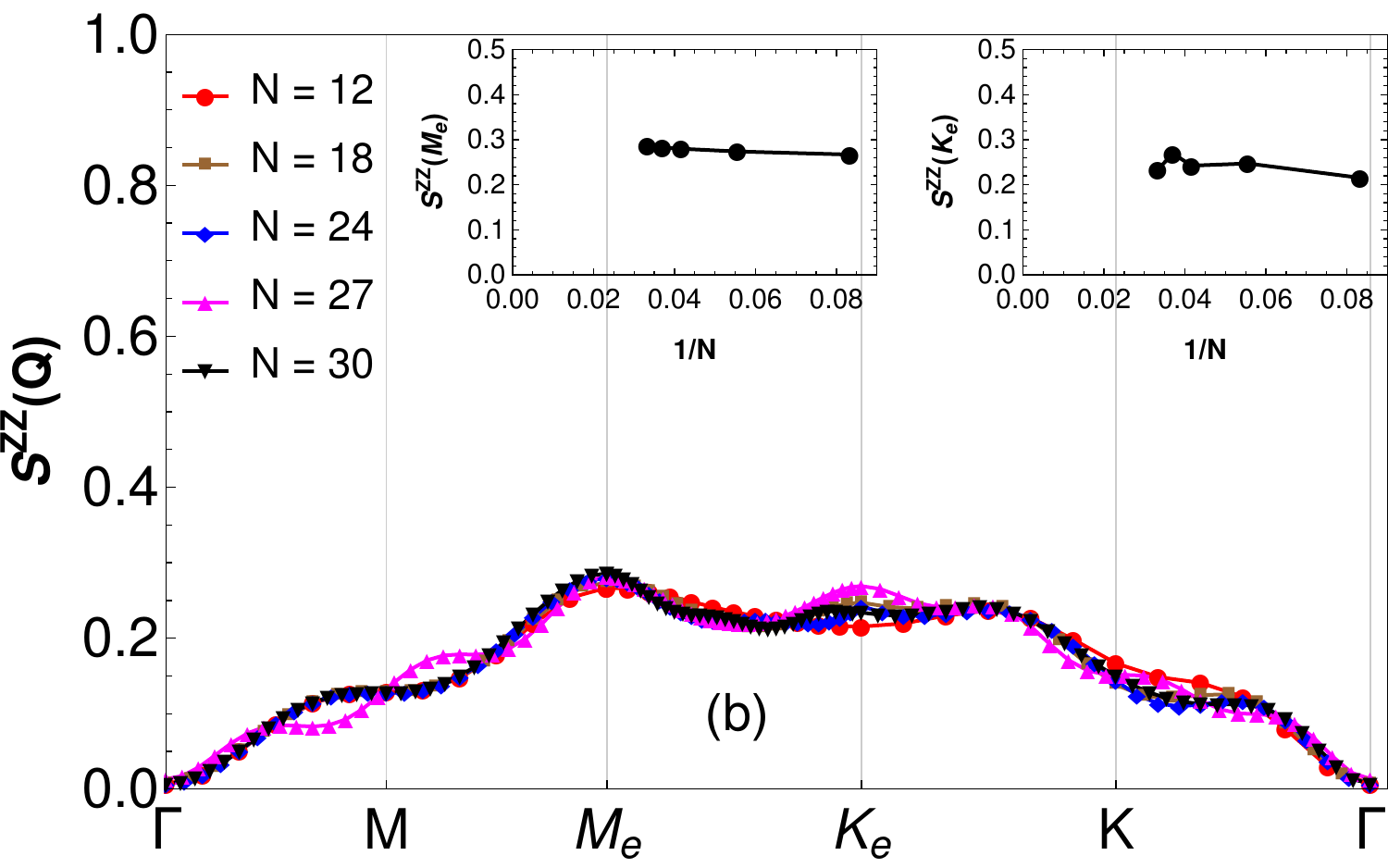}
\end{center}
\caption{(a) XX component of static structure factor (b) ZZ component of static structure factor for  $ Dz = -1$ and $D_p =1$ }
\label{fig:sfdp1dz1}
\end{figure*}
The phase phase diagram is qualitatively similar to the phase diagram obtained from classical calculation as well as SBMFT calculation.
 \begin{figure*}[ht!]
\begin{center}
\includegraphics[height=5.5cm,width=8.25cm]{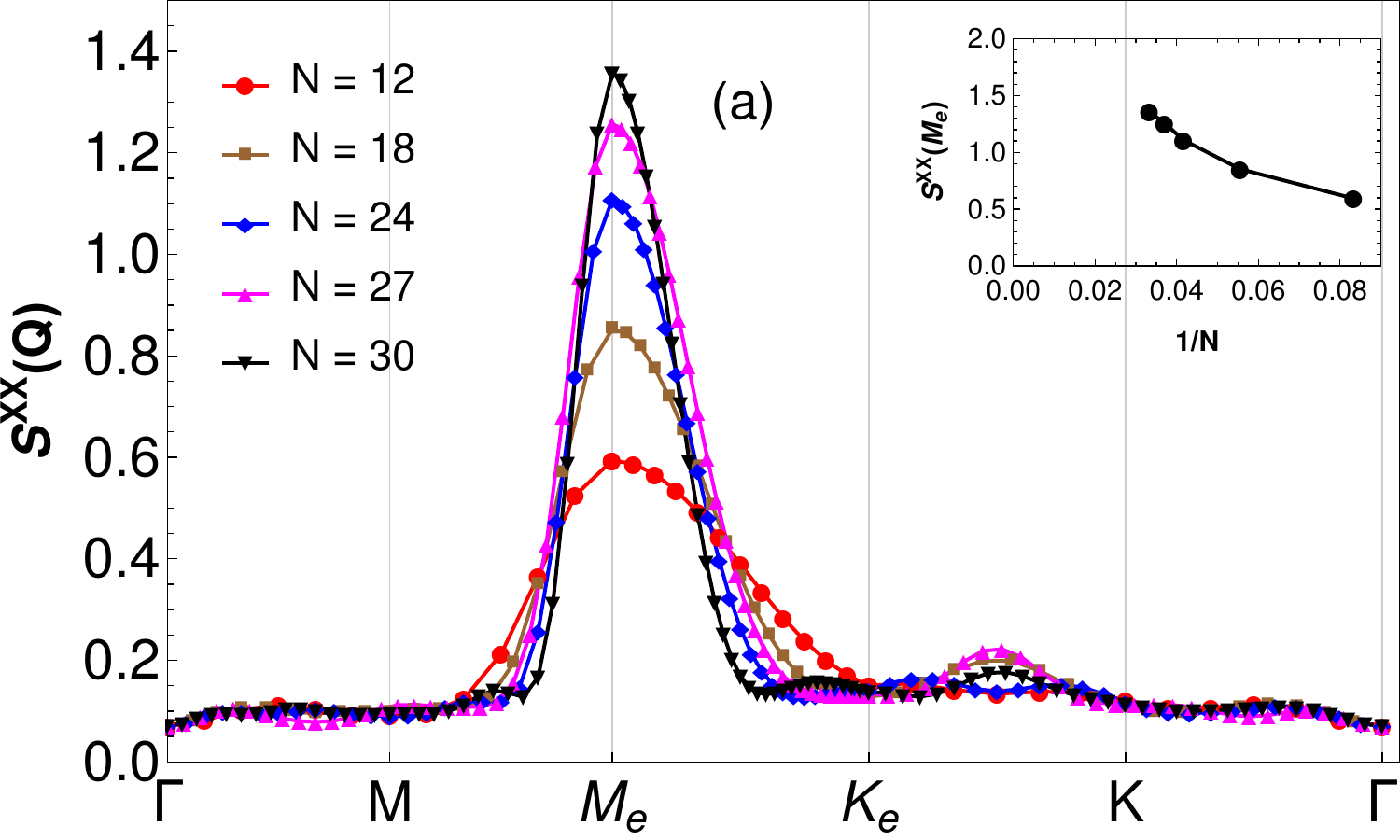}\hspace{1cm}
\includegraphics[height=5.5cm,width=8cm]{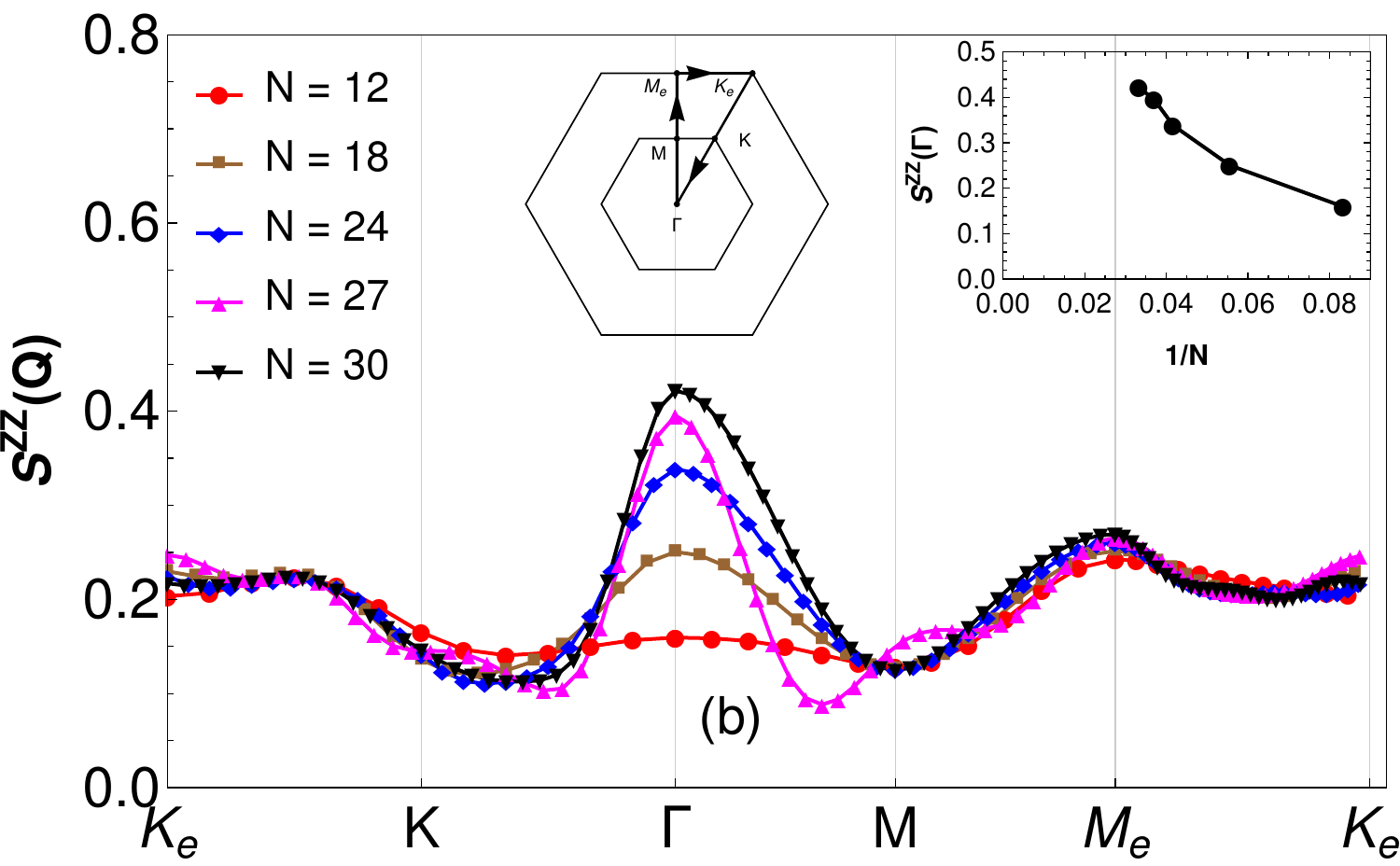}
\end{center}
\caption{ (a) XX component of static structure factor (b) ZZ component of static structure factor for $ D_z =D_p =1$ }
\label{fig:sfdp1dzm1}
\end{figure*} 

In order to establish the ground state spin configuration, we have calculated the static spin structure factor (as defined in Eq.~\ref{eqn:sf}) along the high-symmetry line of the Brillouin zone. The Fig.~\ref{fig:sfdp1dz1} show structure factors for a representative point $D_p = 1 $ and $D_z =- 1$ for the left side of the phase diagram. The XX component of the static structure shows a peak at $M_e$, the height of which diverges in the $N\to\infty$ limit, as shown in the inset of Fig.~\ref{fig:sfdp1dz1}(a). This is a clear sign of the long-range magnetic order of $\mathbf{Q=0}$ type. In this case, the ZZ-component of static structure factor does not show any such peak(see Fig.~\ref{fig:sfdp1dz1}(b)) indicating the planar arrangement of the spins i.e.; the spins lie in the X-Y plane.

In the right part of the phase diagram ($D_z > 0$) we have taken the representative point to be $D_z =1$ and $D_p =1$. Here too, from Fig.~\ref{fig:sfdp1dzm1}(a), we see that the XX component of the static structure shows a peak at $M_e$ showing a long-range magnetic order of $\mathbf{Q=0}$ type. The divergent behavior of the height of the peak is shown in the inset. But the ZZ component of the static structure shows a peak at $\Gamma$ showing a long-range ferromagnetic order, which is increasing as we increase $N$ as shown in the inset in Fig.~\ref{fig:sfdp1dzm1}(b). The canting angle is calculated from this data and is shown in Fig.~\ref{fig:ca}. This shows the umbrella kind of the structure of the ground state.
 
\section{Discussion}
For vesignieite, the measured value of in-plane component $D_p = 0.19J$ and the out-of-plane component $D_z =0.07J$~\cite{PhysRevB.88.144419} which are the two dominant term compared to other anisotropies like isotropic exchange anisotropy. The ground state is expected to be influenced by both in-plane as well as the out-of-plane component of DMI. In the classical limit, any small amount of $D_z$ will force the spin to lie in the kagome plane. In the absence of the in-plane component, the critical value $D_c =0.1J$ predicted by ED result, there is a disordered state at one side and ordered state on the other. So, we expect that the presence of in-plane component $D_p$ will affect this critical value. The presence of the in-plane component of DMI is responsible for the tilting of the spins towards the z-axis. The measured value of the canting angle for vesignieite is found to be $3^{o} < \phi < 9^{o}$, as obtained from NMR data analysis.

In our phase diagram, for the spin-1/2 case, the ground state is in the magnetic LRO state. For positive values of $D_z$, we get the canted magnetic structure, and the canting angle increases with the increase of $D_p$. For $D_p= 0.2$ and $D_z = 0.1$, the estimated value of the canting angle is $6.98^o$, which is very close to the canting angle measured in vesignieite as reported by Zorko et al. ~\cite{PhysRevB.88.144419}.

Since, in the previous SBMFT studies, it was argued that lower values of spin (S = 0.366) is found to be a better description of the spin-1/2 case due to the fact that the constraint $n_i = 2 S$ is not implemented exactly rather implemented as an average~\citep{messio2012kagome}. However, even at $S =0.366$, the ground state is also in the magnetic LRO region for all values of $D_p$ and $D_z$. It seems the canted LRO nature of the ground state of vesignieite is dictated by the presence of DM interaction with a dominant component in the kagome plane.
\section{Conclusion}
We have studied the effect of in-plane and the out-of-plane component of Dzyaloshinskii-Moriya interaction on the ground state of spin-1/2 kagome antiferromagnet using Schwinger Boson mean-field theory as well as numerically using exact diagonalization method up to system size $N=30$. We found two chiralities of $\mathbf{Q=0}$ structure in the phase diagram for the spin-1/2 case in both the approaches. In the case of SBMFT, for the lower values of $S$, the spin liquid phase is sandwiched between the above two chiralities. We also found that this spin-liquid region shrinks to the phase boundary between the two chiralities in the large $S$ limit. In the classical limit, our SBMFT result is in agreement with the result obtained from the classical phase diagram, as well as exact diagonalization results.
\section{Appendix}
\subsection{Proof of Eq.~\ref{eqn:rotated-H}}
Here, we provide the proof for Eq.~\ref{eqn:rotated-H}.  Let us consider the triangle $\Delta \equiv (123) $ as shown in the Fig.~\ref{DM-config-classical-pd}(a). So, the Hamiltonian for this part is given by
\begin{equation}
H_{\Delta} = H_{12} + H_{23} + H_{31}
\end{equation}
where $H_{12}$ has the following form
\begin{equation}
H_{12} = J \vec{\mathbf{S}}_1 \cdot \vec{\mathbf{S}}_2 + \vec{\mathbf{D}}^P_{12} \cdot (\vec{\mathbf{S}}_1 \times \vec{\mathbf{S}}_2) +\vec{\mathbf{D}}^z_{12} \cdot (\vec{\mathbf{S}}_1 \times \vec{\mathbf{S}}_2)
\label{eqn:H12}
\end{equation}
 Similarly, we can write the expressions  for the other two bonds. Now, consider the bond-$12$. We treat each of the spins as a three dimensional classical unit vectors. Then, we rotate the spins by $\theta$ and $- \theta$ around the axis $\hat{\mathbf{d}}_{12}$ where,  $\theta = \frac{\sqrt{D_P^2 +D_z^2 }}{2J}$ and $\hat{\mathbf{d}}_{12}$ is a unit vector along $\vec{\mathbf{D}}_{12}$. For small values of $\theta$, the relation between the rotated and unrotated spins is as given below
\begin{eqnarray}
\vec{\mathbf{S}}_1 & = & \vec{\mathbf{S}}^\prime_1 + \theta~ ( \vec{\mathbf{S}}^\prime_1 \times \hat{\mathbf{d}}_{12})  \\
\vec{\mathbf{S}}_2 & = & \vec{\mathbf{S}}^\prime_2 -  \theta~ ( \vec{\mathbf{S}}^\prime_2 \times \hat{\mathbf{d}}_{12})  
\end{eqnarray}
where,  $\vec{\mathbf{S}}_1^\prime $ and $ \vec{\mathbf{S}}_2^\prime$ be the rotated spins. Then, we can $\hat{\mathbf{d}}_{12}= \frac{1}{2 \theta J}\big[D_P ~\hat{j} + D^z ~\hat{k}\big]$ and neglect $\theta^2$ term to write the Heienberg term as
\begin{equation}
\vec{\mathbf{S}}_1 \cdot \vec{\mathbf{S}}_2 = \vec{\mathbf{S}}_1^\prime \cdot \vec{\mathbf{S}}_2^\prime + 2 ~\theta~ \hat{\mathbf{d}}_{12}\cdot (\vec{\mathbf{S}}_2^\prime \times \vec{\mathbf{S}}_1^\prime)
\label{eqn:rotH}
\end{equation}
The cross product of the DM interaction term in Eq.~\ref{eqn:H12} reduces to 
\begin{eqnarray}
\vec{\mathbf{S}}_1 \times \vec{\mathbf{S}}_2& = & \vec{\mathbf{S}}_1^\prime \times \vec{\mathbf{S}}_2^\prime + \theta~\big[ 2~\hat{\mathbf{d}}_{12}~ (\vec{\mathbf{S}}_1^\prime \cdot \vec{\mathbf{S}}_2^\prime )\nonumber \\
& -  & \vec{\mathbf{S}}_1^\prime~( \hat{\mathbf{d}}_{12}\cdot \vec{\mathbf{S}}_2^\prime ) -  \vec{\mathbf{S}}_2^\prime~( \hat{\mathbf{d}}_{12}\cdot \vec{\mathbf{S}}_1^\prime )  \big]
\label{eqn:rotDM}
\end{eqnarray}
Using Eq.~\ref{eqn:rotH} and Eq.~\ref{eqn:rotDM} in Eq.~\ref{eqn:H12} we get,
\begin{eqnarray}
H_{12} &=& J~ \vec{\mathbf{S}}_1^\prime \cdot \vec{\mathbf{S}}_2^\prime +(2 \theta^2 J \hat{\mathbf{d}}_{12}) \cdot \big[ 2~\hat{\mathbf{d}}_{12}~ (\vec{\mathbf{S}}_1^\prime \cdot \vec{\mathbf{S}}_2^\prime ) \nonumber \\
& -&  \vec{\mathbf{S}}_1^\prime~( \hat{\mathbf{d}}_{12}\cdot \vec{\mathbf{S}}_2^\prime ) -  \vec{\mathbf{S}}_2^\prime~( \hat{\mathbf{d}}_{12}\cdot \vec{\mathbf{S}}_1^\prime )  \big] 
\end{eqnarray}
The second term in the RHS is second order in $\theta$ and hence, can be neglected. So, we finally have
\begin{equation}
H_{12} = J~ \vec{\mathbf{S}}_1^\prime \cdot \vec{\mathbf{S}}_2^\prime
\end{equation}
Similar excercise can be done for bond-23 and bond-31 to get the following form
\begin{equation}
H_{\Delta} = J \big[ \vec{\mathbf{S}}_1^\prime \cdot \vec{\mathbf{S}}_2^\prime +  \vec{\mathbf{S}}_2^{\prime \prime} \cdot \vec{\mathbf{S}}_3^{\prime\prime} +  \vec{\mathbf{S}}_3^{\prime \prime \prime} \cdot \vec{\mathbf{S}}_1^{\prime\prime\prime} \big]
\end{equation}
\subsection{Static structure factor}
The ZZ- component of static structure factor is given by 
\begin{eqnarray}
S^{\text{ZZ}}(\mathbf{Q}) & = & \frac{3}{4 N}\sum_{ij} e^{i \mathbf{Q}\cdot (\mathbf{R}_i - \mathbf{R}_j)} \langle 0 | S^z_i S^z_j | 0 \rangle  \\
&= & \frac{3}{16 N}\sum_{ij} e^{i\mathbf{Q}\cdot (\mathbf{R}_i - \mathbf{R}_j)} \langle 0 | ( [\mathbf{\Phi}^\dagger_i]^T \sigma_z \mathbf{\Phi}_i )\cdot( [\mathbf{\Phi}^\dagger_j]^T \sigma_z \mathbf{\Phi}_j)| 0 \rangle \nonumber 
\end{eqnarray}
After taking the Fourier transform we have,
\begin{eqnarray}
S^{\text{ZZ}}(\mathbf{Q}) & = & \frac{3}{16 N}\sum_{\mu \nu = 1}^{3} \sum_{\mathbf{k} \mathbf{k}^\prime \in BZ}\langle 0 |([\xi^\dagger_{\mathbf{k},\mu}]^T \sigma_z \xi_{\mathbf{Q} +\mathbf{k}, \mu} ) \nonumber \\
&\cdot & ([\xi^\dagger_{\mathbf{Q} + \mathbf{k}^\prime,\nu}]^T \sigma_z \xi_{ \mathbf{k}^\prime, \nu} )| 0 \rangle 
\end{eqnarray}
Let us rearrange the $M$ matrix such that $\Psi^T_\mathbf{k} = \{\xi_{\mathbf{k},1}, \xi_{\mathbf{k},2},\xi_{\mathbf{k},3},\xi^\dagger_{-\mathbf{k},1},\xi^\dagger_{-\mathbf{k},2},\xi^\dagger_{-\mathbf{k},3}\} $. Then we can write 
\begin{equation}
\Psi_\mathbf{k} = M \tilde{\Psi}_\mathbf{k}
\label{appendix:bg}
\end{equation}
 where $M$ has the form
$ M= \begin{pmatrix} 	U  & V \\	X  & Y \end{pmatrix}$ and $\tilde{\Psi}^T_\mathbf{k} = \{\tilde{\xi}_{\mathbf{k},1}, \tilde{\xi}_{\mathbf{k},2},\tilde{\xi}_{\mathbf{k},3}, \tilde{\xi}^\dagger_{-\mathbf{k},1}, \tilde{\xi}^\dagger_{-\mathbf{k},2}, \tilde{\xi}^\dagger_{-\mathbf{k},3}\} $. The reciprocal lattice vectors $\mathbf{k}$ and $\mathbf{k}^\prime$ belongs to the full Brillouin zone formed by $\mathbf{k}_a$ and $\mathbf{k}_b$, where $\mathbf{k}_a =(\sqrt{3},-1)2\pi/\sqrt{3} $ and $\mathbf{k}_b = (4 \pi/\sqrt{3})(0,1)$. From Eq.~\ref{appendix:bg}, we can write,
\begin{eqnarray}
\xi_{\mathbf{k},\mu} &  = & \sum^3_{\rho=1}  \Big[[A_\mathbf{k}]_{\mu \rho} \tilde{\xi}_{\mathbf{k},\rho} + [B_\mathbf{k}]_{\mu \rho}\tilde{\xi}^\dagger_{-\mathbf{k},\rho} \Big] 
\end{eqnarray} 
where  $A$ and $B$ can be substituted by $U, X $ or $V, Y$. Now $S^{\text{ZZ}}(\mathbf{Q})$ can be written as a sum of two terms, 
\begin{eqnarray}
 S^{\text{ZZ}}(\mathbf{Q})  & = &  S^{\text{ZZ}}(\mathbf{Q\neq 0}) +  S^{\text{ZZ}}(\mathbf{Q=0}) 
\end{eqnarray}
where
\begin{eqnarray}
S^{\text{ZZ}}(\mathbf{Q\neq 0}) & = & \frac{3}{16 N} \sum_{\mathbf{k}} \sum_{\mu \nu} \sum_{\alpha \beta}\big[\text{Tr}[[B_\mathbf{k}]^{\dagger}_{\mu \alpha} [A^\prime_{\mathbf{Q}+\mathbf{k}}]_{\mu \beta}\cdot \nonumber \\
&  &  ([A_{-\mathbf{k}}]^{\dagger}_{\nu \alpha} [B^\prime_{-(\mathbf{Q}+\mathbf{k})}]_{\nu \beta}])^T +  \text{Tr}[[B_\mathbf{k}]^{\dagger}_{\mu \beta} [A^\prime_{\mathbf{Q}+\mathbf{k}}]_{\mu \alpha}\cdot\nonumber  \\
&  & [A_{\mathbf{Q}+\mathbf{k}}]^{\dagger}_{\nu \alpha} [B^\prime_{\mathbf{k}}]_{\nu \beta}] \big]  \\
S^{\text{ZZ}}(\mathbf{Q= 0}) & = & \frac{3}{16 N}\sum_{\mu \alpha }\sum_{\mathbf{k}} \text{Tr}[ [B_\mathbf{k}]^{\dagger}_{\mu \alpha}[B^\prime_{\mathbf{k}}]_{\mu \alpha}] \nonumber \\
&\times& \sum_{ \nu \beta } \sum_{\mathbf{k}^\prime}  \text{Tr}[[B_{\mathbf{k}^\prime}]^{\dagger}_{\nu \beta} [B^\prime_{\mathbf{k}^\prime}]_{\nu \beta}] 
\end{eqnarray}
For the XX- component of structure factor, $\sigma_z$ will be replaced by $\sigma_x$. 
\bibliography{referance}
\bibliographystyle{apsrev4-1}
\end{document}